\documentclass[namedreferences]{solarphysics}
\usepackage[optionalrh]{spr-sola-addons} 
\usepackage{graphicx}        
\usepackage{amssymb}        
\usepackage[usenames,dvipsnames]{color}           
\usepackage{url}             

\newcommand{\aap}{    {\it Astron. Astrophys.}}

\newcommand{\apj}{    {\it Astrophys. J.}}

\newcommand{\jgr}{    {\it J. Geophys. Res.}}

\newcommand{\solphys}{{\it Solar Phys.}}

\begin{document}

\begin{article}

\begin{opening}

\title{Statistical Survey of Type III Radio Bursts at Long Wavelengths Observed by the \textit{Solar TErrestrial RElations Observatory} (STEREO)/\textit{Waves} Instruments: Goniopolarimetric Properties and Radio Source Locations}

\author{V.~\surname{Krupar}$^{1}$\sep
        M.~\surname{Maksimovic}$^{2}$\sep  
        O.~\surname{Santolik}$^{1,3}$\sep
        B.~\surname{Cecconi}$^{2}$\sep
        O.~\surname{Kruparova}$^{1}$ 
       }
\runningauthor{Krupar et al.}
\runningtitle{Type III Radio Bursts}

   \institute{$^{1}$ Institute of Atmospheric Physics ASCR, Bocni II 1403, Prague 141 31, Czech Republic
                     email: \url{vk@ufa.cas.cz}\\ 
              $^{2}$ LESIA, UMR CNRS 8109, Observatoire de Paris, Meudon 92195, France
                      \\ 
              $^{3}$ Faculty of Mathematics and Physics, Charles University, Ke Karlovu 3, Prague 121 16, Czech Republic
             }

\begin{abstract}
We have performed statistical analysis of a large number of Type III
radio bursts observed by STEREO between May 2007 and February 2013. Only intense, simple, and isolated cases have been included
in our data set.
We have focused on the goniopolarimetric (GP, also referred to as direction-finding) properties
at frequencies between $125$~kHz and $2$~MHz.
The apparent source size $\gamma$ is very extended ($\approx60^\circ$) for the lowest analyzed frequencies.
Observed apparent source sizes $\gamma$  expand linearly with a radial distance from the Sun at frequencies below $1$~MHz.
We have shown that Type III radio bursts statistically propagate in the ecliptic plane.
Calculated positions of radio sources suggest
that scattering of the primary beam pattern plays an important role in the propagation of Type III radio bursts in the IP medium.

\end{abstract}
\end{opening}

\section{Introduction}
     \label{S-introduction}
Solar radio emissions represent a clear evidence of far-reaching effects of solar eruptions injecting energetic electrons into the interplanetary (IP) medium.
Kinetic energy of these electrons can be partly converted into electromagnetic radiation.
Therefore solar radio emissions provide us with an important diagnostic tool for remote measurements of the electron density and magnetic field between the Sun and the Earth.
These emissions are often called radio bursts to emphasize their brief and explosive characteristics.

Type III radio bursts belong among the most intense electromagnetic emissions observed in the heliosphere \citep{1950AuSRA...3..541W,1987A&A...173..366D}.
They are consequences of impulsively accelerated electrons associated with solar flares. The mechanism of energy release in solar flares is believed to be related to the reconnection of the magnetic field lines.
The magnetic field creates the current sheet representing a very efficient particle accelerator \citep{1989SoPh..122..263M}.
The downward moving electrons radiate in the X-ray spectra, whereas the electrons propagating outward from the Sun along open magnetic field lines can
produce an intense radio emission: Type III radio bursts.

Type III-generating electron beams produce a bump-on-tail instability of the electron velocity distribution function \citep{1958SvA.....2..653G,1980SSRv...26....3M}.
This abrupt change in plasma parameters trigger electrostatic Langmuir waves at the local
plasma frequency $f_{\rm{pe}}$, which is proportional to the square root of the local electron density
($f_{\rm{pe}}\rm{[kHz]} \approx 9 $$\sqrt{n_{\rm{e}}\rm{[cm^{-3}]}}$).
Langmuir waves can be afterward converted by nonlinear interactions into Type III radio bursts \citep{1976Sci...194.1159G,1981ApJ...251..364L}.
Langmuir waves scattered off of ions or low-frequency turbulence result in electromagnetic waves at $f_{\rm{pe}}$ (the fundamental emission).
Two Langmuir waves can coalesce to produce radiation at $2f_{\rm{pe}}$ (the harmonic emission).
This generation mechanism is often referred to as the plasma emission mechanism.
Since electrons propagate outward from the Sun to lower densities, these emissions are generated at lower
frequencies corresponding to a decreasing in $f_{\rm{pe}}$.
Therefore the locations of Type III radio bursts enable us to track
solar energetic particles and to map the global magnetic field topology in the corona and IP medium.
Measurements of IP Type III radio bursts from ground-based observatories
are restricted by the ionospheric cut-off frequency and radio instruments
in space are thus needed despite their high requirements on design.

Radio sources can be tracked with one spacecraft with goniopolarimetric
(GP, also referred to as direction-finding) capabilities using an average IP density model
\citep{1972Sci...178..743F}. However, the IP density profile along the path of exciter electrons
is not obvious at the given time and location.
In a case of long-lasting Type III storm we may localize radio sources using solar rotation \citep{1984A&A...141...17B}.
Multi-spacecraft observations allow to triangulate radio sources without any additional assumptions on the IP density model
\citep{1977SoPh...54..431W,1977SoPh...52..477F,1998JGR...103.1923R,2009SoPh..tmp..100R}.
The \textit{Solar TErrestrial RElations Observatory} (STEREO)
is the first three-axis stabilized solar mission that consists of two spacecraft embarking identical radio receivers
which enable us to accurately track source locations of solar radio emissions \citep{2008SSRv..136....5K,2008SSRv..136..549C}.

%

This is the second of two linked papers that summarize our findings on statistical properties of Type III radio bursts
observed by the STEREO/\textit{Waves} instruments.
In this paper we focus on GP properties.
The frequency spectra and its relation to electron beams producing these bursts have been discussed in the first paper
\citep{2014SoPh..289.3121K}. In Section \ref{S-data} we describe source data set and methods used. In Section \ref{S-stat}
we present a detailed analysis of two Type III radio bursts as an example of our data set.
We discuss statistical results of a large number of Type III radio bursts.
In Section \ref{S-conclusion} we summarize our findings and we make concluding remarks.

\section{Instrumentation and Methodology} 
      \label{S-data}

\subsection{The STEREO/\textit{Waves} Instrument} 
      \label{S-waves}
The STEREO/\textit{Waves} instrument provides comprehensive measurements of the three
components of the  electric field fluctuations up to 16 MHz \citep{2008SSRv..136..487B}.
Three mutually orthogonal monopole antenna elements
(each six meters in length) form the sensor part of the STEREO/\textit{Waves} instrument \citep{2008SSRv..136..529B}.
The High Frequency Receiver (HFR, a part of STEREO/\textit{Waves})
has instantaneous GP capabilities between $125$~kHz and $2$~MHz encompassing both the direction-finding and polarization \citep{2008SSRv..136..549C}.

The STEREO/\textit{Waves} effective antenna directions
and lengths are different from the physical ones due to their electric coupling with
the spacecraft body.
The galactic background, as a nearly stable isotropic source, allowed us to determine reduced effective antenna lengths \citep{2011RaSc...46.2008Z}.
The effective antenna directions have been obtained by observations of the auroral
kilometric radiation (AKR) during STEREO-B roll maneuvers \citep{2012JGRA..11706101K}.
As no AKR has been surveyed by STEREO-A, the effective directions have been assumed to be the same.

\subsection{Singular Value Decomposition of the Electric Spectral Matrix}
The singular value decomposition (SVD) is an iterative method for a wave analysis of
multi-component measurements of the magnetic field with a point source \citep{2003RaSc...38a..10S}.
We may extend its application to electric field measurements
when a wave frequency $f$ is significantly larger than the local electron plasma frequency $f_{\rm{pe}}$,
which is the case of radio waves observed by STEREO/\textit{Waves}/HFR \citep{2010AIPC.1216..284K,2012JGRA..11706101K}.

Applying SVD on the spectral matrix $\mathbf{A}$ composed from electric complex amplitudes $\hat{E}_k$
with separated real and imaginary parts we obtain matrices $\mathbf{U}$, $\mathbf{W}$, and
$\mathbf{V}^{\rm{T}}$:
\begin{equation}
\mathbf{A}=\mathbf{U}\cdot\mathbf{W}\cdot\mathbf{V}^{\rm{T}}\rm{.}
\label{A_UWV2}
\end{equation}
The diagonal nonnegative matrix $\mathbf{W}$ in Equation (\ref{A_UWV2}) contains three singular values that represent relative lengths of the axes of the polarization ellipsoid in ascending order: $w_1$, $w_2$, and $w_3$ are minimal, medial, and maximal singular values.
These singular values reveal electric field variances along axes of the polarization ellipsoid.
In a case of a monochromatic plane wave with a point source the polarization ellipsoid degenerates to an ellipse ($w_1/w_3 = 0$).
On the other hand if no direction is preferred ($w_1=w_2=w_3$), the polarization ellipsoid degenerates and becomes a sphere.
Therefore we assume that information about the angular half aperture of the source ($\gamma$) as seen by the spacecraft
is hidden in the ratio $w_1/w_3$.
We have used an empirical relation between $w_1/w_3$ and $\gamma$ considering unpolarized emissions and the Gaussian radial source shape
\citep{2012JGRA..11706101K}.

SVD provides us with three characteristic directions of the polarization ellipsoid in the rows of the $\mathbf{V}^{\rm{T}}$ matrix:
the wave vector, minor polarization axis, and major polarization axis directions. These three directions are mutually orthogonal.
The wave vector direction $\mbox{\boldmath$\kappa$}=\mathbf{k}/|\mathbf{k}|$ is in the row that corresponds to $w_1$, \textit{i.e.} to the shortest axis of 
the polarization ellipsoid.

The degree of polarization describes a quantity of an electromagnetic wave which is polarized.
We use the two-dimensional (2D) degree of polarization in the polarization plane ($C$) according to \citet{2002JGRA..107.1444S}:
\begin{equation}
C=\sqrt{2\frac{R_{33}^2+R_{22}^2+2|R_{23}|}{R_{33}+R_{22}}-1}\rm{,}
\label{C_R}
\end{equation}
where $R_{ij}$ is the spectral matrix $A_{ij}$ transformed into a matrix in the primary axial system (\textit{i.e.} the polarization plane):
\begin{eqnarray}
R_{kl} = \sum\limits_{m=1}^3 W_{ll} U_{ml} V_{mk} \mbox{ for } k=1,2,3 \mbox{ and } l = 1,2,3,\\
R_{kl} = \sum\limits_{m=4}^6 W_{ll} U_{ml} V_{m-3 k-3} \mbox{ for } k=4,5,6 \mbox{ and } l = 1,2,3 \mbox{.}
\label{R_def}
\end{eqnarray}

In this paper for a localization of radio sources we use the heliocentric earth equatorial (HEEQ) coordinate system with the following definition:
the $Z_{\rm{HEEQ}}$ axis denotes the solar rotation axis, and the $X_{\rm{HEEQ}}$ axis is in the plane containing the $Z_{\rm{HEEQ}}$ axis and Earth.
For relative wave directions we use the radial-tangential-normal (RTN) coordinate system.
The $X_{\rm{RTN}}$ axis points from the spacecraft to the Sun's center, and the $Y_{\rm{RTN}}$ axis is the cross product
of the solar rotational axis and $X_{\rm{RTN}}$ lying in the solar equatorial plane towards the West limb.

\subsection{Triangulation}
\label{S-triang}
\begin{figure}    
   \centerline{\includegraphics[width=0.8\textwidth,clip=]{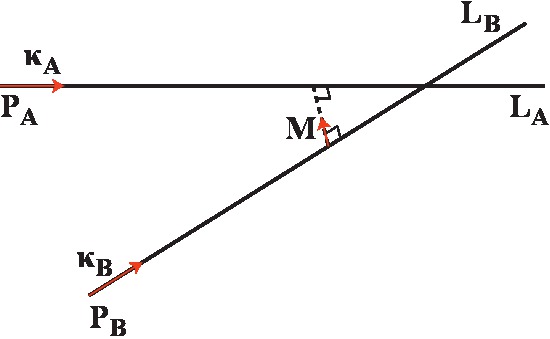}
              }
              \caption{Illustration of a triangulation of radio sources. $\mathbf{P}_{\rm{A}}$ and $\mathbf{P}_{\rm{B}}$ are spacecraft
							positions. $\mbox{\boldmath$\kappa$}_{\rm{A}}$ and $\mbox{\boldmath$\kappa$}_{\rm{B}}$ represent wave vector directions.
							These parameters form two straight lines: $\mathbf{L}_{\rm{A}}$ and $\mathbf{L}_{\rm{B}}$. The $\mathbf{M}$ is
							a cross product of $\mbox{\boldmath$\kappa$}_{\rm{A}}$ and $\mbox{\boldmath$\kappa$}_{\rm{B}}$.}
   \label{triang}
   \end{figure}

Aforementioned determination of the wave vector direction $\mbox{\boldmath$\kappa$}$ allows us to estimate
a source location if an electromagnetic emission is observed by two (or more) spacecraft.
This process is referred to as the triangulation.
The source location can be found as an intersection of lines $\mathbf{L}_{\rm{A}}$ and $\mathbf{L}_{\rm{B}}$ in three dimensional (3D) space (Figure \ref{triang}):
\begin{eqnarray}
\mathbf{L}_{\rm{A}}=\mathbf{P}_{\rm{A}}+t_{\rm{A}}\boldmath{\mbox{\boldmath$\kappa$}}_{\rm{A}}\mbox{,}
\label{eq:triang_a}
\end{eqnarray}
\begin{eqnarray}
\mathbf{L}_{\rm{B}}=\mathbf{P}_{\rm{B}}+t_{\rm{B}}\boldmath{\mbox{\boldmath$\kappa$}}_{\rm{B}}\mbox{,}
\label{eq:triang_b}
\end{eqnarray}
where $\mathbf{P}_k$ denotes spacecraft position vector, $\mbox{\boldmath$\kappa$}_k$
represents a wave vector direction and $t_k$ is a parameter we need to retrieve.
Two lines intersect in 3D at one point only if they lie in the same plane.
In a general case, we have to determine the closest point on line A ($\mathbf{r}_{\rm{A}}$) and the closest point on line B ($\mathbf{r}_{\rm{B}}$).
First we calculate the direction between two spacecraft ($\mathbf{P}_{\rm{BA}}$) and the direction perpendicular to both wave
vector directions ($\mathbf{M}$):
\begin{eqnarray}
\mathbf{P}_{\rm{BA}}=\mathbf{P}_{\rm{B}}-\mathbf{P}_{\rm{A}}\mbox{,}\\
\mathbf{M}=\boldmath{\mbox{\boldmath$\kappa$}}_{\rm{B}}\times\boldmath{\mbox{\boldmath$\kappa$}}_{\rm{A}}\mbox{.}
\label{eq:triang_1}
\end{eqnarray}
Then we calculate the direction $\mathbf{D}$ which is perpendicular to $\mathbf{P_{\rm{BA}}}$ and $\mathbf{M}$:
\begin{equation}
\mathbf{D}=\frac{\mathbf{P}_{\rm{BA}}\times\mathbf{M}}{||\mathbf{M}||^2}\mbox{.}
\label{eq:r_dir}
\end{equation}
From $\mathbf{D}$ we can derive parameters $t_{\rm{AI}}$ and $t_{\rm{BI}}$ as its projections into the mutual wave vector directions:
\begin{eqnarray}
t_{\rm{AI}}=\mathbf{D}\cdot\boldmath{\mbox{\boldmath$\kappa$}}_{\rm{B}}\mbox{,}\\
t_{\rm{BI}}=\mathbf{D}\cdot\boldmath{\mbox{\boldmath$\kappa$}}_{\rm{A}}\mbox{.}
\label{eq:triang_2}
\end{eqnarray}
Finally, we obtain the closest points on each of lines:
\begin{eqnarray}
\mathbf{r}_{\rm{A}}=\mathbf{P}_{\rm{A}}+t_{\rm{AI}}\boldmath{\mbox{\boldmath$\kappa$}}_{\rm{A}}\mbox{,}\\
\mathbf{r}_{\rm{B}}=\mathbf{P}_{\rm{B}}+t_{\rm{BI}}\boldmath{\mbox{\boldmath$\kappa$}}_{\rm{B}}\mbox{.}
\label{eq:triang_3}
\end{eqnarray}
In this paper we consider the intersection ($\mathbf{r}$) to be the closest point between two lines
which can be calculated as a mean distance between $\mathbf{r}_{\rm{A}}$ and $\mathbf{r}_{\rm{B}}$:
\begin{equation}
\mathbf{r}=\frac{\mathbf{r}_{\rm{A}}+\mathbf{r}_{\rm{B}}}{2}\mbox{.}
\label{eq:triang_mean}
\end{equation}
Projections of a segment line $\mathbf{r}_{\rm{A}}$ and $\mathbf{r}_{\rm{B}}$ reveal an accuracy of the triangulation ($e_{\rm{i}}$) in the given direction:
\begin{equation}
e_{\rm{i}}=|r_{\rm{Ai}}-r_{\rm{Bi}}|
\label{eq:err}
\end{equation}

\section{Results and Discussion} 
      \label{S-stat}

We have manually selected 152 time-frequency intervals when Type III radio bursts have been observed by STEREO/\textit{Waves} between May 2007
and February 2013.
The separation angle between the spacecraft in the ecliptic plane ranged between $7^\circ$ (May 2007) and $180^\circ$ (February 2011).
We have included only simple and isolated events when flux density $S$ was intense enough for the GP analysis.
We have linked
$20$\,\% of Type III radio bursts with solar flares using database compiled
by the Lockheed Martin Solar and Astrophysics Laboratory (\url{http://www.lmsal.com/solarsoft/latest_events_archive.html}).
A detailed description of the data set can be found in the first article \citep{2014SoPh..289.3121K}.
As an example from our list of events we present two Type III radio bursts observed in
January 2008 and in May 2009.


\subsection{29 January 2008 Type III Burst}
\label{S-case1}

\begin{figure}    
   \centerline{\includegraphics[width=0.8\textwidth,clip=]{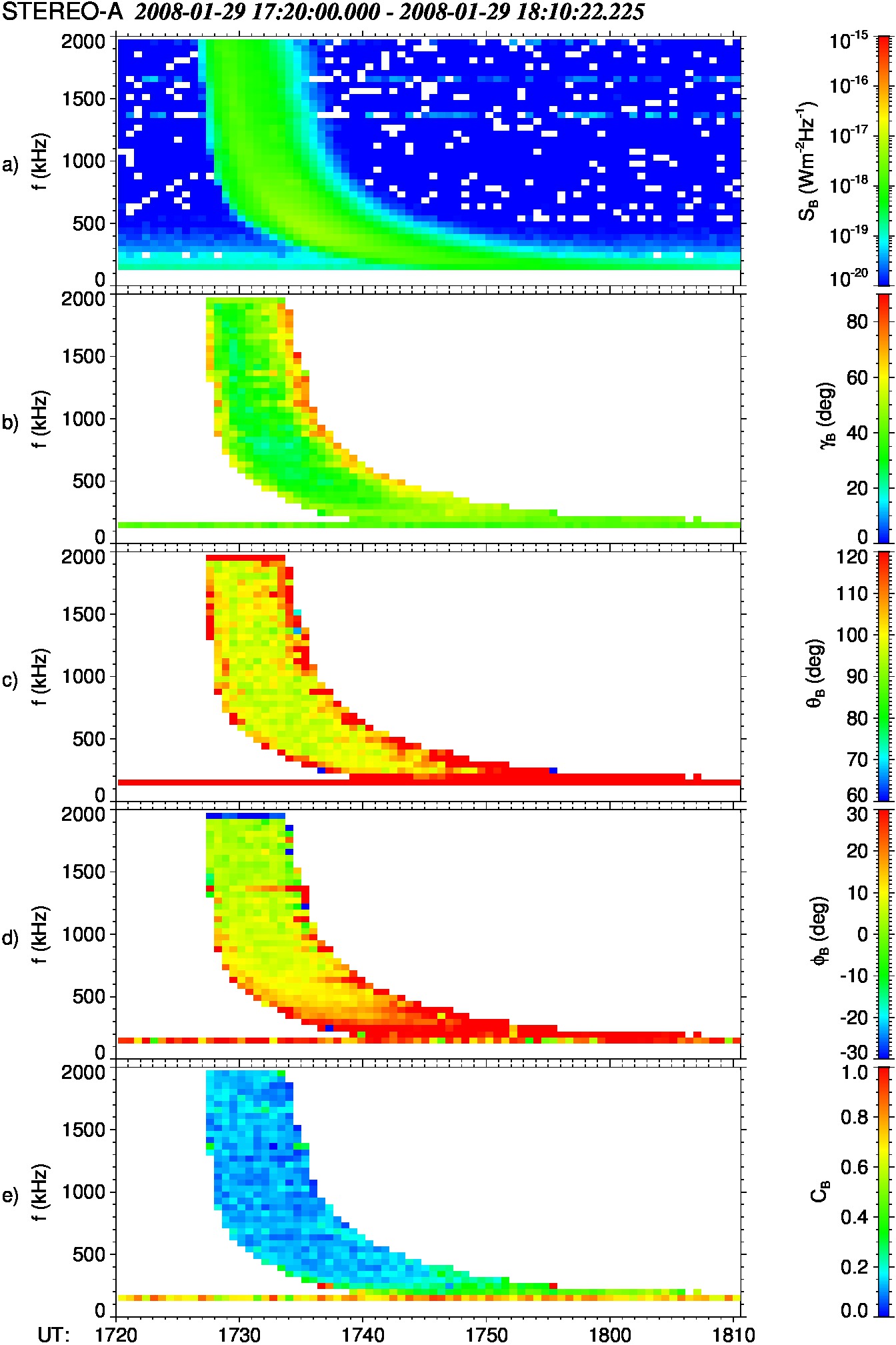}
              }
              \caption{From 17:20 to 18:10 UT on 28 January 2008: a) flux density $S_{\rm{A}}$,
							b) the apparent source size $\gamma_{\rm{A}}$, c) the colatitude $\theta_{\rm{A}}$ in RTN, d) the azimuth $\phi_{\rm{A}}$ in RTN, and
							e) the 2D degree of polarization $C_{\rm{A}}$  for STEREO-A.
							The intensity threshold ($S_{\rm{A}}>10^{-19}$~$\rm{Wm^{-2}Hz^{-1}}$) has been applied on the  b), c), d) and e) panels to suppress the background.}
   \label{20090501a_spectra}
   \end{figure}

\begin{figure}    
   \centerline{\includegraphics[width=0.8\textwidth,clip=]{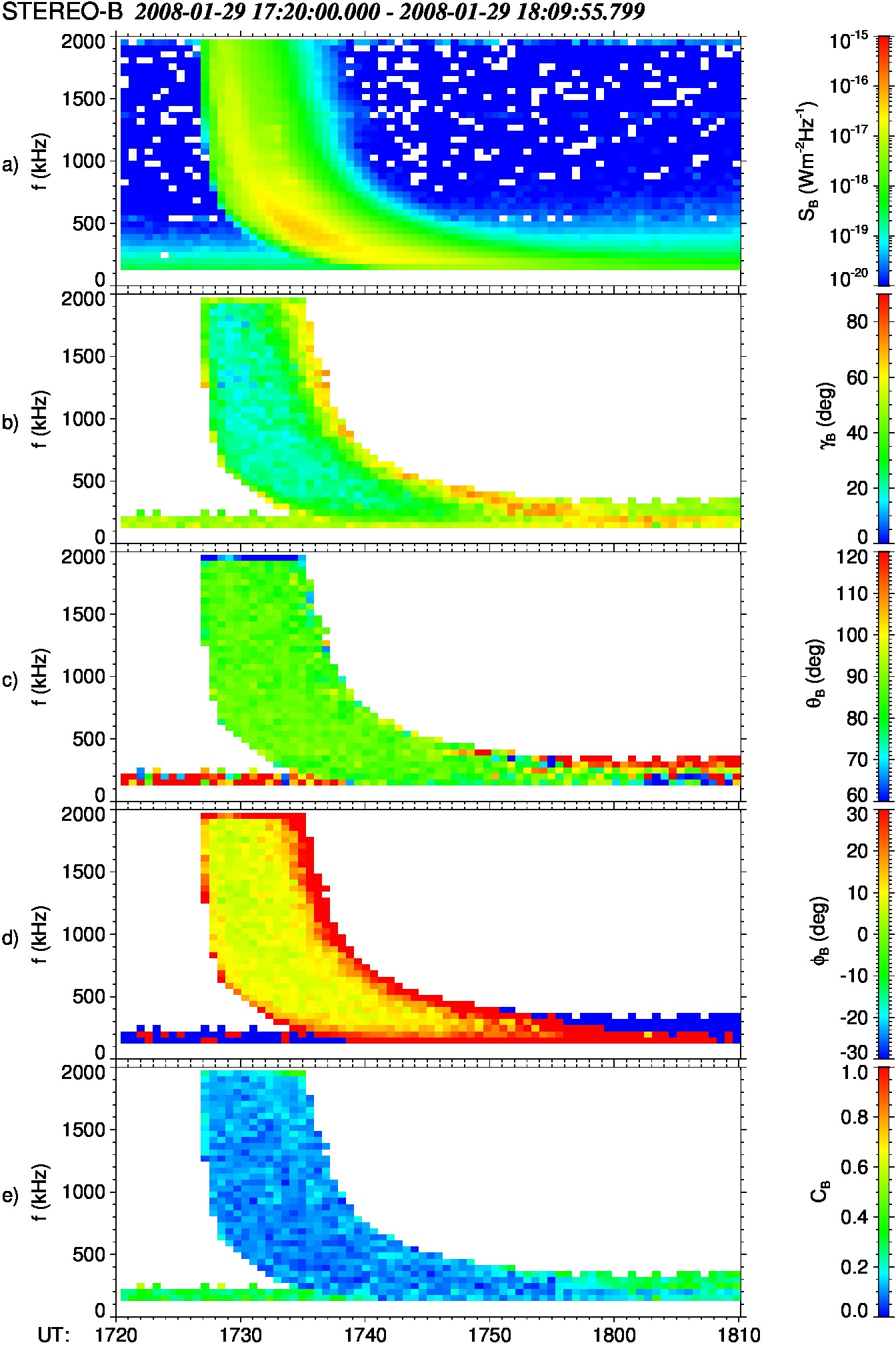}
              }
              \caption{From 17:20 to 18:10 UT on 28 January 2008: a) flux density $S_{\rm{B}}$,
							b) the apparent source size $\gamma_{\rm{B}}$, c) the colatitude $\theta_{\rm{B}}$, d) the azimuth $\phi_{\rm{B}}$, and
							e) the 2D degree of polarization $C_{\rm{B}}$ for STEREO-B.
							The intensity threshold ($S_{\rm{B}}>10^{-19}$~$\rm{Wm^{-2}Hz^{-1}}$) has been applied on the  b), c), d) and e) panels to suppress the background.}
   \label{20090501b_spectra}

   \end{figure}

We show an observation of a Type III radio burst from
29 January 2008, which has been already analyzed by \citet{2009SoPh..tmp..100R}
and \citet{2012SoPh..279..153M}.  It has been linked to a B1.2 solar flare
starting at 17:28 UT reaching maximum at 17:43 UT.
The flare site was at S$09^\circ$E$59^\circ$ in the HEEQ coordinate system.
During this event STEREO-A was located $21.6^\circ$
west from a Sun-Earth line at $0.97$~AU
from the Sun whereas STEREO-B was at $23.6^\circ$ east and $1.00$~AU from the Sun.
Figures \ref{20090501a_spectra} and \ref{20090501b_spectra} show the data recorded on STEREO-A and STEREO-B, respectively.
Both STEREO detected a simple and isolated Type III radio burst with a starting time of about 17:28~UT.
The signal measured by STEREO-B was significantly larger
than by STEREO-A
which suggests that radio source is located closer to STEREO-B.

\begin{figure}    
   \centerline{\includegraphics[width=0.8\textwidth,clip=,angle=90]{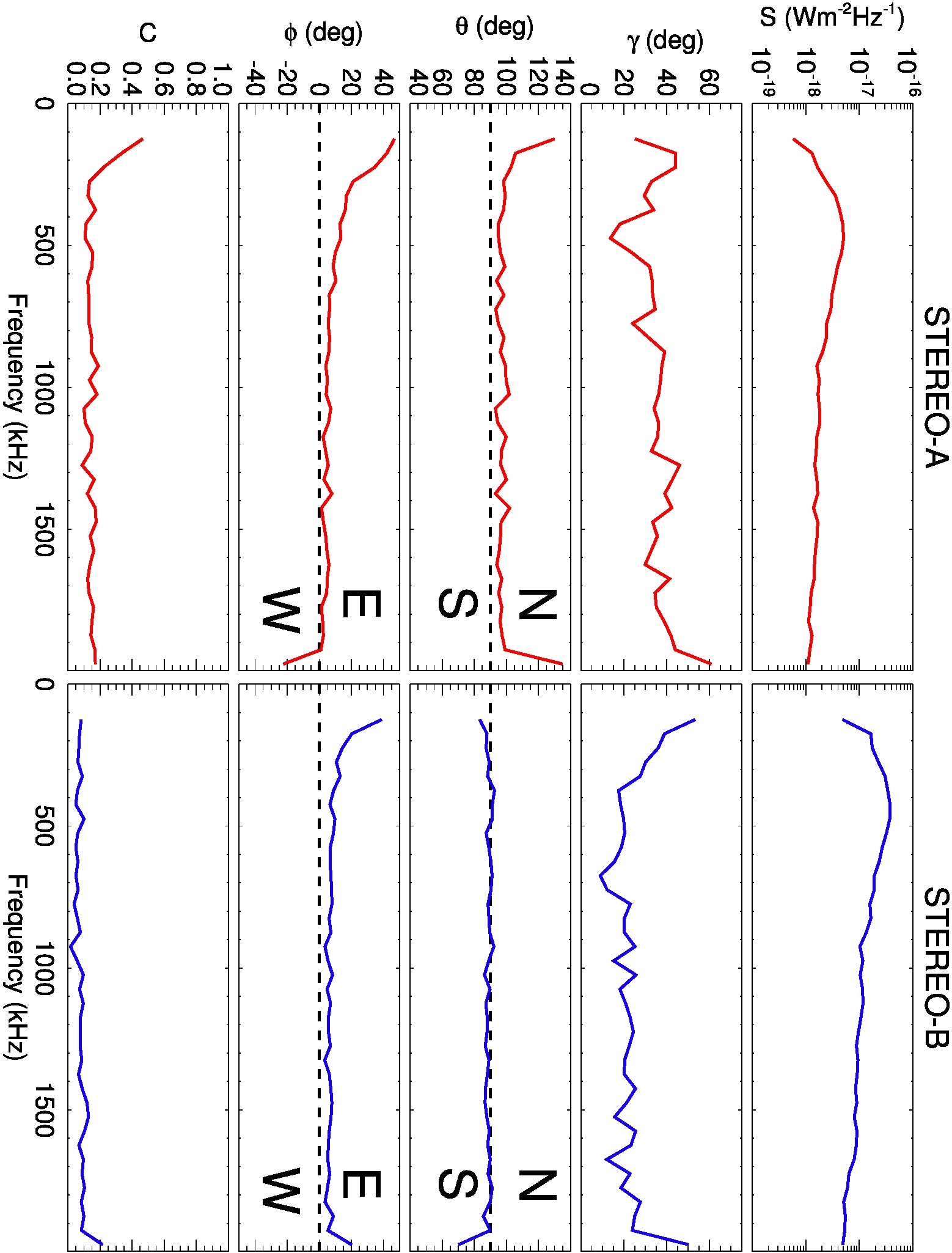}
              }
              \caption{From 17:20 to 18:10 UT on 28 January 2008: the peak values of the flux density $S$,
							the colatitude $\theta$, the azimuth $\phi$,
							the 2D degree of polarization $C$, and the apparent source size $\gamma$
							during peak fluxes vs frequency for STEREO-A (on the left) and STEREO-B (on the right).}
   \label{20090501_ls}
   \end{figure}

We have identified data points corresponding to peak fluxes for each frequency separately (Figure \ref{20090501_ls}).
The flux density $S_{\rm{B}}$ is about of one order of magnitude larger than $S_{\rm{A}}$
with the maximum at $475$~kHz at both spacecraft. The flux density $S_{\rm{A}}$ for the lowest frequency channel is
very weak ($\approx 5\times10^{-19}$~$\rm{Wm^{-2}Hz^{-1}}$) being comparable with background.
The apparent source sizes $\gamma_{\rm{A}}$ and $\gamma_{\rm{B}}$ are very extended at both spacecraft.
While $\gamma_{\rm{B}}$ is largest for the lowest frequency channels corresponding to previous studies \citep{1985A&A...150..205S},
$\gamma_{\rm{A}}$ exhibits no frequency dependence. It can be explained by weak $S_{\rm{A}}$ when the
planarity of the polarization, used for $\gamma$ estimation, is dominated by background signals.
The wave vector directions $\theta_{\rm{A}}$/$\theta_{\rm{B}}$ (the Sun direction: $\theta=90^\circ$, the southward direction: $\theta=0^\circ$,
and the northward direction: $\theta=180^\circ$) and $\phi_{\rm{A}}$/$\phi_{\rm{B}}$ (the Sun direction: $\phi=0^\circ$, the eastward direction: $\phi=90^\circ$,
and the westward direction: $\phi=-90^\circ$) point roughly towards the Sun.
Only the low frequency part of $\phi_{\rm{A}}$/$\phi_{\rm{B}}$ shows an eastward shift which corresponds to the flare site.
The highest frequency channel (1975~kHz) is significantly affected by instrumental
effects when HFR switches from the three monopoles mode (HFR1, GP data) to the dipole/monopole mode (HFR2, no GP data).
The last panel of Figure \ref{20090501_ls} contains the 2D degree of polarization
which is about $0.2$ at STEREO-A and $0.1$ at STEREO-B that
is comparable with limits of an accuracy of the receiver itself. The increase at $125$~kHz at STEREO-A is probably related to the weak signal.

\begin{figure}    
   \centerline{\includegraphics[angle=90,width=0.85\textwidth]{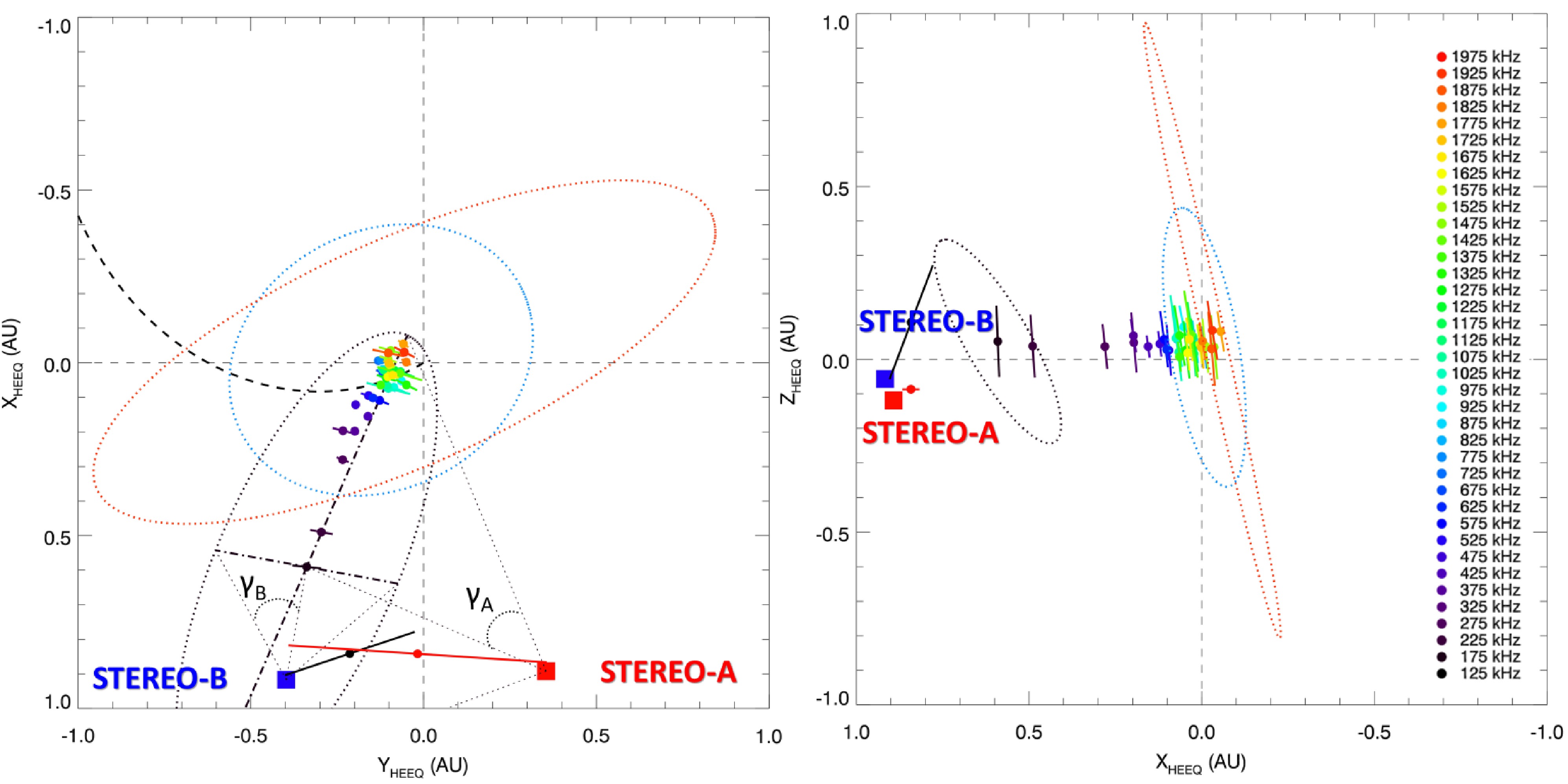}
              }
              \caption{Propagation analysis of measurements recorded from 17:20 to 18:10 UT on 28 January 2008:
							Circles show intersections between wave vector directions (peak flux values) from STEREO-A and STEREO-B
							in the $XY_{\rm{HEEQ}}$ and $XZ_{\rm{HEEQ}}$ planes. Colors denote frequency.
							Segment lines indicate accuracy of the triangulation.
							Dotted ellipses around three intersections ($175$~kHz, $1075$~kHz, and $1925$ kHz) represent apparent source sizes as seen from both spacecraft.
							Dot-dashed lines illustrate how the ellipse around the $175$~kHz intersection has been calculated.
							The black dashed line shows the Parker spiral.
							Positions of STEREO-A and STEREO-B are denoted by red and blue squares, respectively. The Sun is located at the center.}
   \label{intersections1s}
   \end{figure}

We have performed a triangulation of radio sources
(Section \ref{S-triang}) using wave vector directions during peak fluxes (Figure \ref{intersections1s}).
The accuracy of radio source locations calculated from Equation~(15) is significantly larger in the $XY_{\rm{{HEEQ}}}$ plane than in the $XZ_{\rm{{HEEQ}}}$ plane
due to the smaller separation angle between the two STEREO in the latter.
We have also included the Parker spiral rooted in the flare site (S$09^\circ$E$59^\circ$) assuming
a typical solar wind speed of $500$~$\rm{km\mbox{ }s^{-1}}$ to illustrate the path followed by the electron beam \citep{1958ApJ...128..664P}.
We have achieved a good agreement between the Parker
spiral and radio locations for frequencies above $700$~kHz. We suggest that the low frequency part is
shifted due to propagation effects in the IP medium \citep{1984A&A...140...39S,2007ApJ...671..894T}.
The dotted ellipses around three intersections ($175$~kHz, $1075$~kHz, and $1925$~kHz) represent the apparent source sizes as seen from both spacecraft.
The two dashed-dotted segment lines show linear source sizes were obtained from a source location, $\gamma_A$, and $\gamma_B$ for $175$~kHz.
The ellipse around these segment lines is the smaller one from the only two possible solutions.
We have included the $1975$~kHz frequency channel (located at $[0.8,0.0,-0.1]_{\rm{HEEQ}}$) to illustrate the aforementioned instrumental
effects of HFR switching.
The error bar on the $125$~kHz frequency channel (located at $[0.8,-0.2,0.1]_{\rm{HEEQ}}$) indicates that the obtained radio source location is unreliable.

Results of the triangulation confirm an assumption that
suprathermal electrons triggering the Type III burst propagate in the ecliptic \citep{1986ASSL..123..229D}.
Source regions of higher frequencies are located closer to the Sun as it can be expected.
We conclude that calculated radio sources
are located closer to STEREO-B being in agreement
with location of the solar flare site (S$09^\circ$E$59^\circ$), and
detected lower signal at STEREO-A.
Our results on radio source locations
are consistent with previous studies of the same event \citep{2009SoPh..tmp..100R,2012SoPh..279..153M}.

\subsection{2 May 2009 Type III Burst}
\label{S-case2}

\begin{figure}    
   \centerline{\includegraphics[width=0.8\textwidth,clip=]{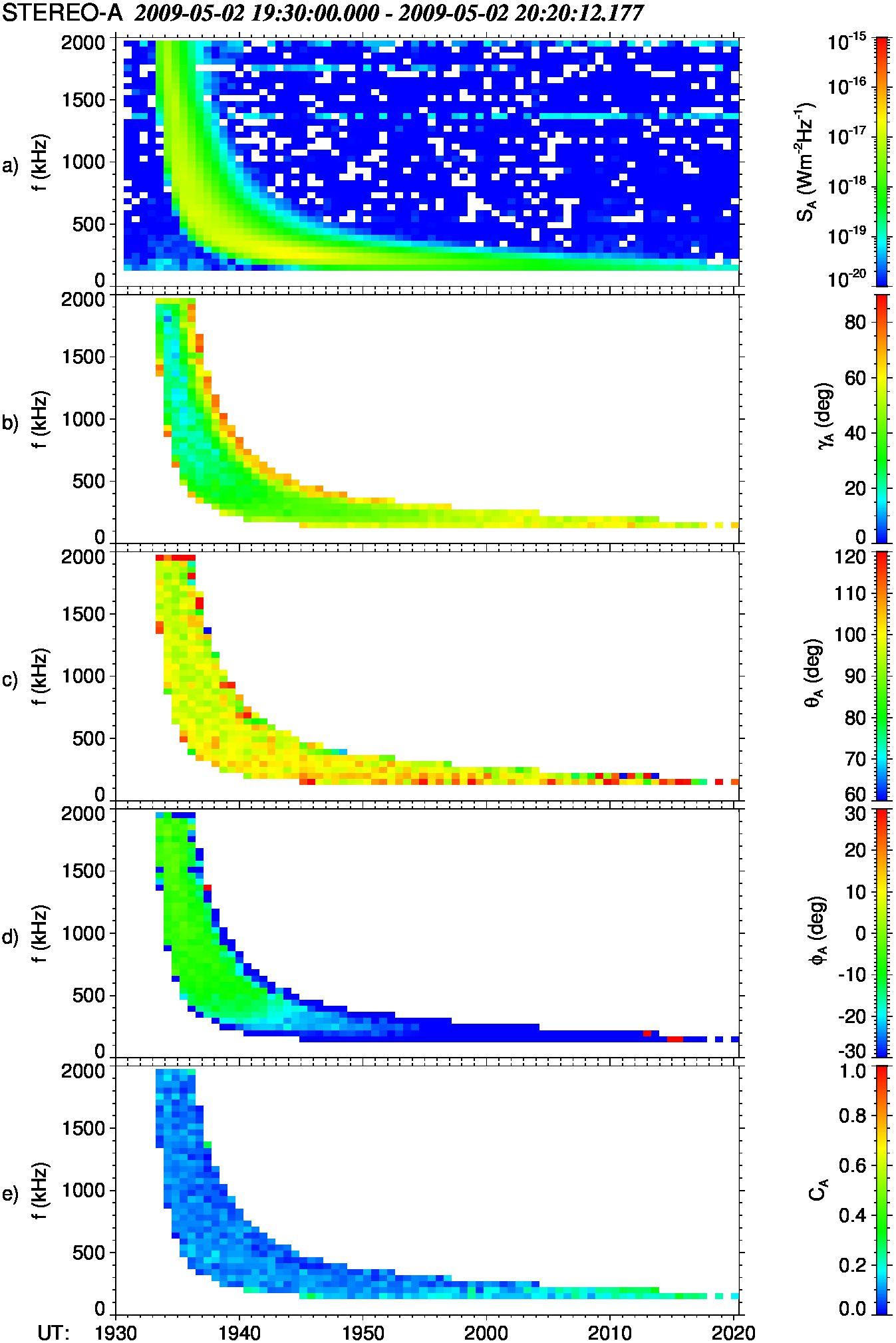}
              }
              \caption{From 19:30 to 20:20 UT on 2 May 2009: a) the flux density $S_{\rm{A}}$,
							b) the apparent source size $\gamma_{\rm{A}}$, c) the colatitude $\theta_{\rm{A}}$, d) the azimuth $\phi_{\rm{A}}$, and
							e) the 2D degree of polarization $C_{\rm{A}}$ for STEREO-A.
							The intensity threshold ($S_{\rm{A}}>10^{-19}$~$\rm{Wm^{-2}Hz^{-1}}$) has been applied on the  b), c), d) and e) panels to suppress the background.}
   \label{20090502a_spectra}
   \end{figure}

\begin{figure}    
   \centerline{\includegraphics[width=0.8\textwidth,clip=]{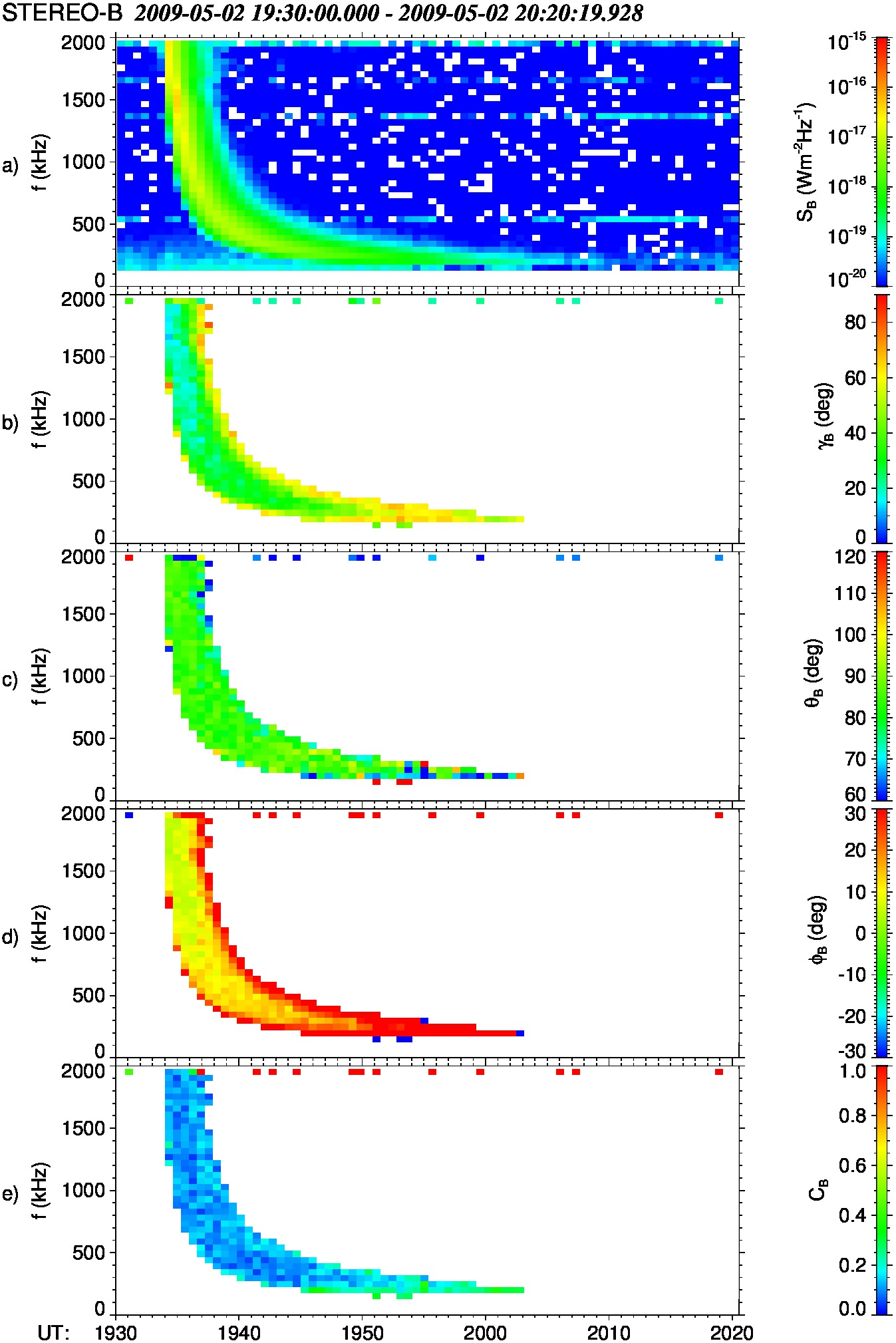}
              }
              \caption{From 19:30 to 20:20 UT on 2 May 2009: a) flux density $S_{\rm{B}}$,
							b) the apparent source size $\gamma_{\rm{B}}$, c) the colatitude $\theta_{\rm{B}}$, d) the azimuth angle $\phi_{\rm{B}}$, and
							e) the 2D degree of polarization $C_{\rm{B}}$ for STEREO-B.
							The intensity threshold ($S_{\rm{B}}>10^{-19}$~$\rm{Wm^{-2}Hz^{-1}}$) has been applied on the  b), c), d) and e) panels to suppress the background.}
   \label{20090502b_spectra}
   \end{figure}

This Type III radio burst occurred on 2 May 2009 at around 19:30 UT. 
STEREO-A was at $47.8^\circ$ west from a Sun-Earth line at $0.95$~AU from the Sun while STEREO-B was located $46.9^\circ$ east at $1.02$~AU from the Sun.
Both STEREO detected a simple and isolated Type III radio burst with a starting time of about 19:30 UT (Figures
\ref{20090502a_spectra} and \ref{20090502b_spectra}).
The same onset times suggest that radio sources are located roughly between the two STEREO.
A solar flare triggering this emission has been located on the far side of the Sun from a view of the Earth.
Hence we cannot retrieve its intensity and location as spacecraft embarking X-ray imagers orbit the Earth.

\begin{figure}    
   \centerline{\includegraphics[width=0.8\textwidth,clip=,angle=90]{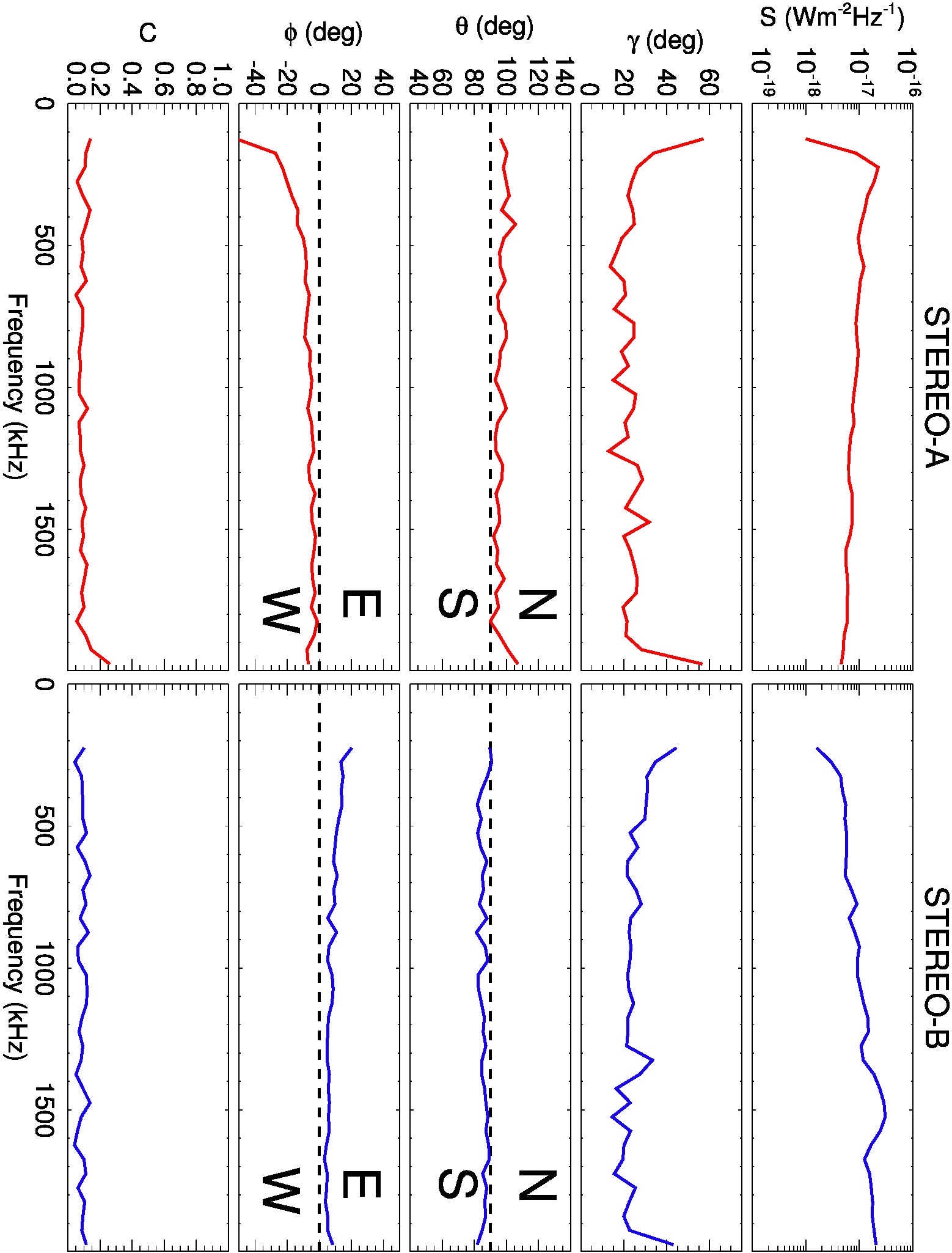}
              }
              \caption{From 19:30 to 20:20 UT on 2 May 2009: peak values of the flux density $S$,
							the colatitude $\theta$ in RTN, the azimuth $\phi$ in RTN,
							the 2D degree of polarization $C$, and the apparent source size $\gamma$
							during peak fluxes vs frequency for STEREO-A (on the left) and STEREO-B (on the right).}
   \label{20090502_ls}
   \end{figure}

We have selected data points that correspond to peak fluxes for each frequency separately (Figure \ref{20090502_ls}).
The flux densities $S_{\rm{A}}$ and $S_{\rm{B}}$ are comparable at both spacecraft while their maxima occur
at $225$~kHz and $1525$~kHz at STEREO-A and STEREO-B, respectively.
The apparent source sizes $\gamma_{\rm{A}}$ and $\gamma_{\rm{B}}$ are very extended with maximum of $60^\circ$ at the lowest
frequency channel.
The wave vector directions $\theta_{\rm{A}}$/$\theta_{\rm{B}}$ and $\phi_{\rm{A}}$/$\phi_{\rm{B}}$ point roughly towards the Sun
for high frequencies whereas the low frequency part of $\phi_{\rm{A}}$ and $\phi_{\rm{B}}$ shows the westward and eastward directions, respectively.
The last panel of Figure \ref{20090502_ls} show the 2D degree of polarization
which is about $0.1$ which is negligible.

\begin{figure}    
   \centerline{\includegraphics[angle=0,width=0.85\textwidth]{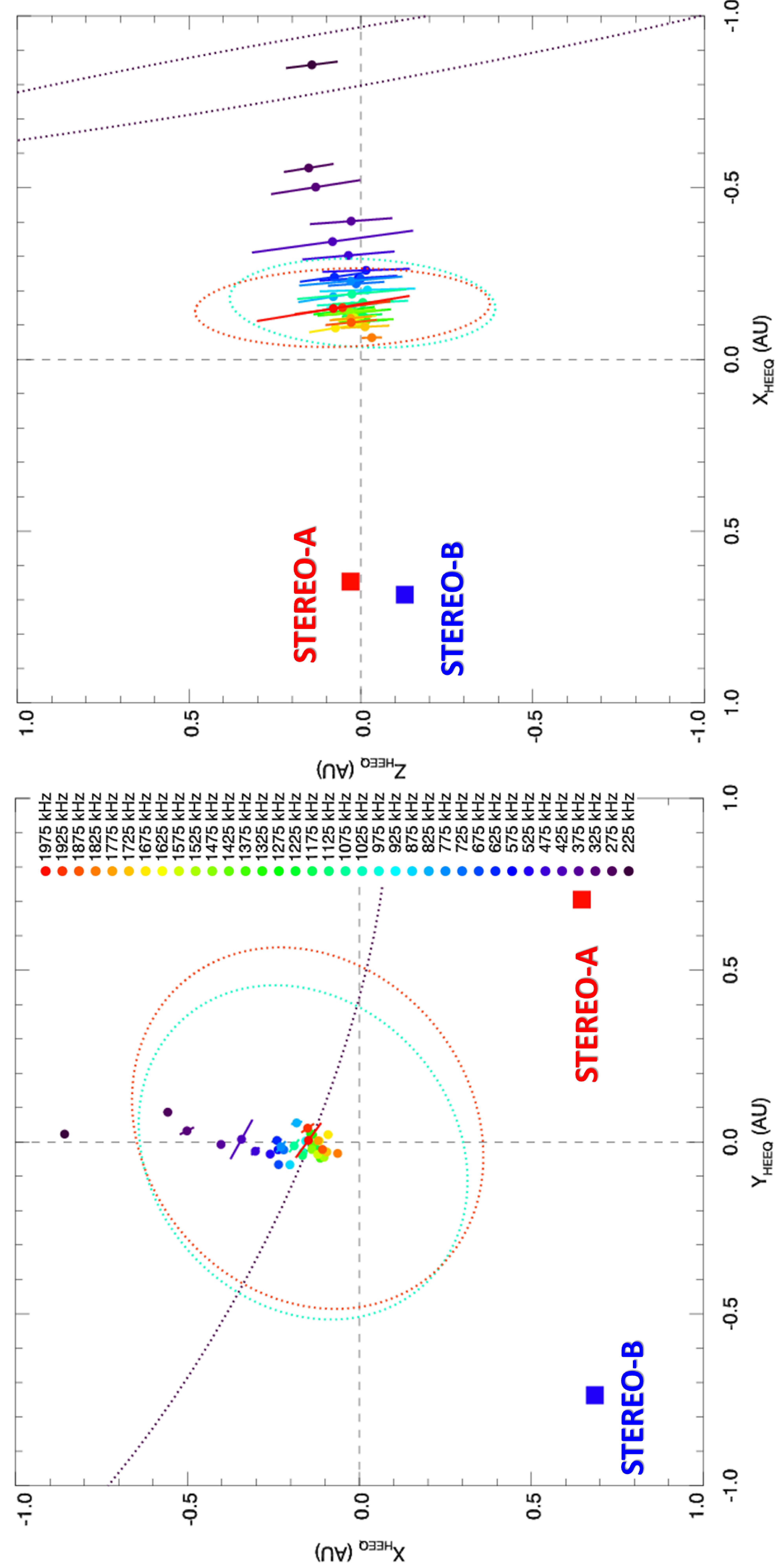}
              }
              \caption{Propagation analysis of measurements recorded from 19:30 to 20:20 UT on 2 May 2009:
							Crosses show intersections between wave vector directions (peak flux values) from STEREO-A and STEREO-B
							in the $XY_{\rm{HEEQ}}$ and $XZ_{\rm{HEEQ}}$ planes. Colors denote frequency.
							Segment lines indicate accuracy of the triangulation.
							Dotted ellipses around three intersections ($225$~kHz, $1075$~kHz, and $1875$ kHz) represent apparent source sizes as seen from both spacecraft.
							Positions of STEREO-A and STEREO-B are denoted by red and blue squares, respectively. The Sun is located at the center.}
   \label{intersections2s}
   \end{figure}

Figure \ref{intersections2s} shows results
of the radio triangulation between $125$~kHz and $1975$~kHz using wave vector
directions during peak fluxes in the $XY_{\rm{HEEQ}}$ and $XZ_{\rm{HEEQ}}$ planes.
Generally, source regions of higher frequencies are closer to the Sun.
Results of the triangulation suggest that an electron beam triggering the Type III burst
propagate along Parker spiral field lines on the far side of the Sun.
We demonstrated a capability of STEREO/\textit{Waves} to track radio sources located up to $1.5$~AU from the spacecraft.

\subsection{Statistical Results on GP Properties}
\label{S-stat1}

\begin{figure}    
   \centerline{\includegraphics[width=1.\textwidth]{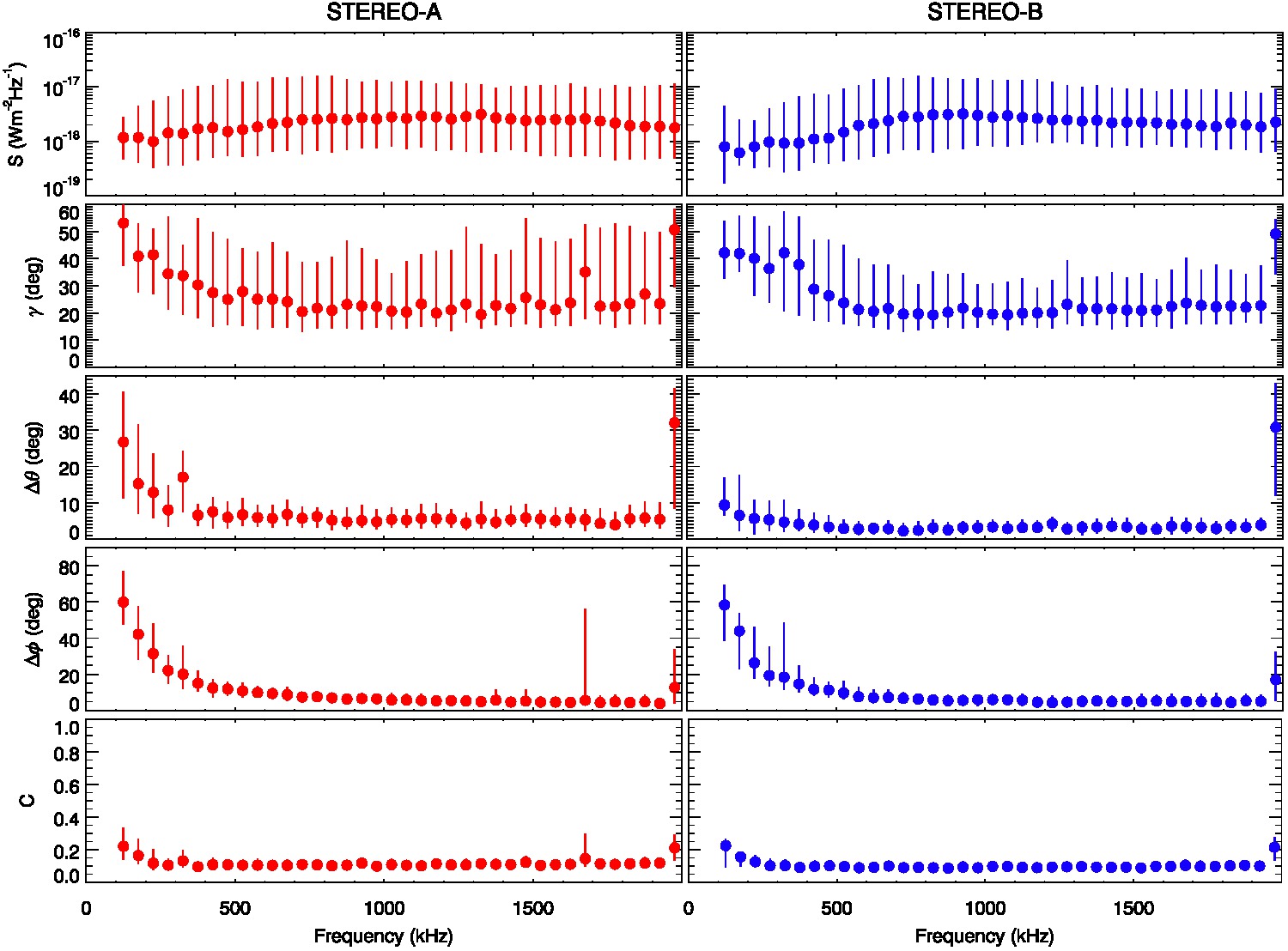}
              }
              \caption{Results of the statistical survey of 152 Type III radio bursts obtained during peak fluxes
							for STEREO-A (on the left) and STEREO-B (on the right).
              The topmost panels display the flux density $S$.
							The second panels show the apparent source sizes $\gamma$.
							The following panels are absolute values of deviations
							of the wave vector from the Sun-spacecraft line out of the ecliptic (the colatitude $\theta$)
							and in the ecliptic (the azimuth $\phi$).
							The bottom panels contain 2D degree of polarization $C$. 
							Circles are median values and errorbars represent 25th/75th percentiles.}
   \label{stat}
   \end{figure}

We have performed a statistical analysis of $152$~events when one or two STEREO spacecraft observed simple and isolated
Type III radio bursts. 
We have investigated median values of the flux density, wave vector directions, and polarization state.
We have selected these data products during peak fluxes case by case for each frequency channel separately.
The topmost panel of Figure \ref{stat} shows median values of the flux density $S$ \textit{vs.} frequency. Both spacecraft observed a
broad maximum at $1$~MHz
being consistent with previous observations using spin-stabilized spacecraft
\citep{1978SoPh...59..377W,1984A&A...141...30D}.
This frequency corresponds to a region ($\approx10R_{\odot}$), where the corona boundary is roughly located and
the radial plasma density gradient changes according to the electron density model \citep{1999ApJ...523..812S}.
However, the observed maximum is shifted in the two events presented in Sections~3.1 and 3.2 with a factor of two and four, respectively. More
discussion on the $1$~MHz maximum can be found in the linked paper \citep{2014SoPh..289.3121K}.

The second panel of Figure \ref{stat} displays median values of the apparent source size
$\gamma$ \textit{vs.} frequency. We have used the GP inversion assuming unpolarized emissions and the Gaussian radial source shape \citep{2012JGRA..11706101K}.
The apparent source size is very extended ($\gamma \approx 60^\circ$) for low frequencies
whereas it is roughly constant ($\gamma \approx 20^\circ$) between $1$~MHz and $2$~MHz
at both spacecraft. The highest frequency channel ($1975$~kHz) is affected by the HFR switching between two modes (Section \ref{S-case1}) and therefore
it should not be taken into account.
Our results for low frequencies are comparable with \citet{1985A&A...150..205S}, who
analyzed $162$~Type III radio bursts observed by the \textit{International Sun/Earth Explorer} 3 (ISEE-3) and concluded that
statistically $\gamma\approx50^\circ$ for $f=100$~kHz.
However, they reported that $\gamma\approx5^\circ$ for $1$~MHz, which is about four-times smaller than we observe by STEREO/\textit{Waves}.
This discrepancy can be explained by different methods for estimating the source size.
In the case of the spin-stabilized ISEE-3 spacecraft, the source size is deduced from the depth of modulation of the
spinning dipole \citep{1980SSI.....5..161M,1981JGR....86.4531H}.
For STEREO/\textit{Waves}, we apply SVD on auto- and cross-spectra measured by three non-orthogonal antennas.
The SVD analysis above $1$~MHz is perhaps distorted by background signals
resulting in increased source sizes. However, it was demonstrated for lower frequencies that the source size calculated by SVD on STEREO is about the same
as on the \textit{Wind} spacecraft (GP capabilities up to $1040$~kHz) during short separation distances using the method dedicated for
spinning spacecraft \citep{2012JGRA..11706101K}.

Statistical results on absolute values of the 
differences between wave vectors and Sun \--- spacecraft lines
out of ($|\theta|$) and in ($|\phi|$) the ecliptic are shown in the third and fourth
panels of Figure \ref{stat}, respectively. Generally, statistics on absolute value represent
deviations of a given value from mean values (the Sun \--- spacecraft line in this case).
These deviations for the lowest frequencies are significantly larger
in the ecliptic plane than in the directions perpendicular to it.
They occur to be twice larger for STEREO-A ($|\theta|_{\rm{A}} \approx 30^\circ$ \textit{vs.} $|\phi|_{\rm{A}} \approx 60^\circ$), whereas
this difference for STEREO-B is about a factor of four ($|\theta|_{\rm{B}} \approx 40^\circ$ \textit{vs.} $|\phi|_{\rm{B}} \approx 10^\circ$).
We do not have any explanation for this discrepancy yet.
It might be caused by slightly different effective antenna parameters between the two spacecraft.
We note that used effective directions have been retrieved by the inflight calibration for STEREO-B only and applied to
STEREO-A too.
However, distributions of $|\theta|$ and $|\phi|$ suggest that electrons triggering
Type III bursts statistically propagate preferentially in the ecliptic \citep{1986ASSL..123..229D}.

The last panel of Figure \ref{stat} is the median value of the 2D degree of polarization in the polarization plane $C$
which is very low ($\approx0.1$). It is in an agreement with previous observations
at kilometric wavelengths when Type III radio bursts
have very low degree of polarization when compared to ground-based measurements \citep{1980A&A....88..203D}.
A slight increase of $C$ at low frequencies ($<200$~kHz) 
is probably induced by a larger error of the wave vector direction calculation
due to extremely large apparent source sizes as the source position and its polarization are determined simultaneously
\citep{2012JGRA..11706101K}.
The low values of $C$ are probably linked to the receiver noise.
Therefore we conclude that observed Type III radio bursts are nearly unpolarized.

\begin{figure}    
   \centerline{\includegraphics[width=0.8\textwidth,angle=0,clip=]{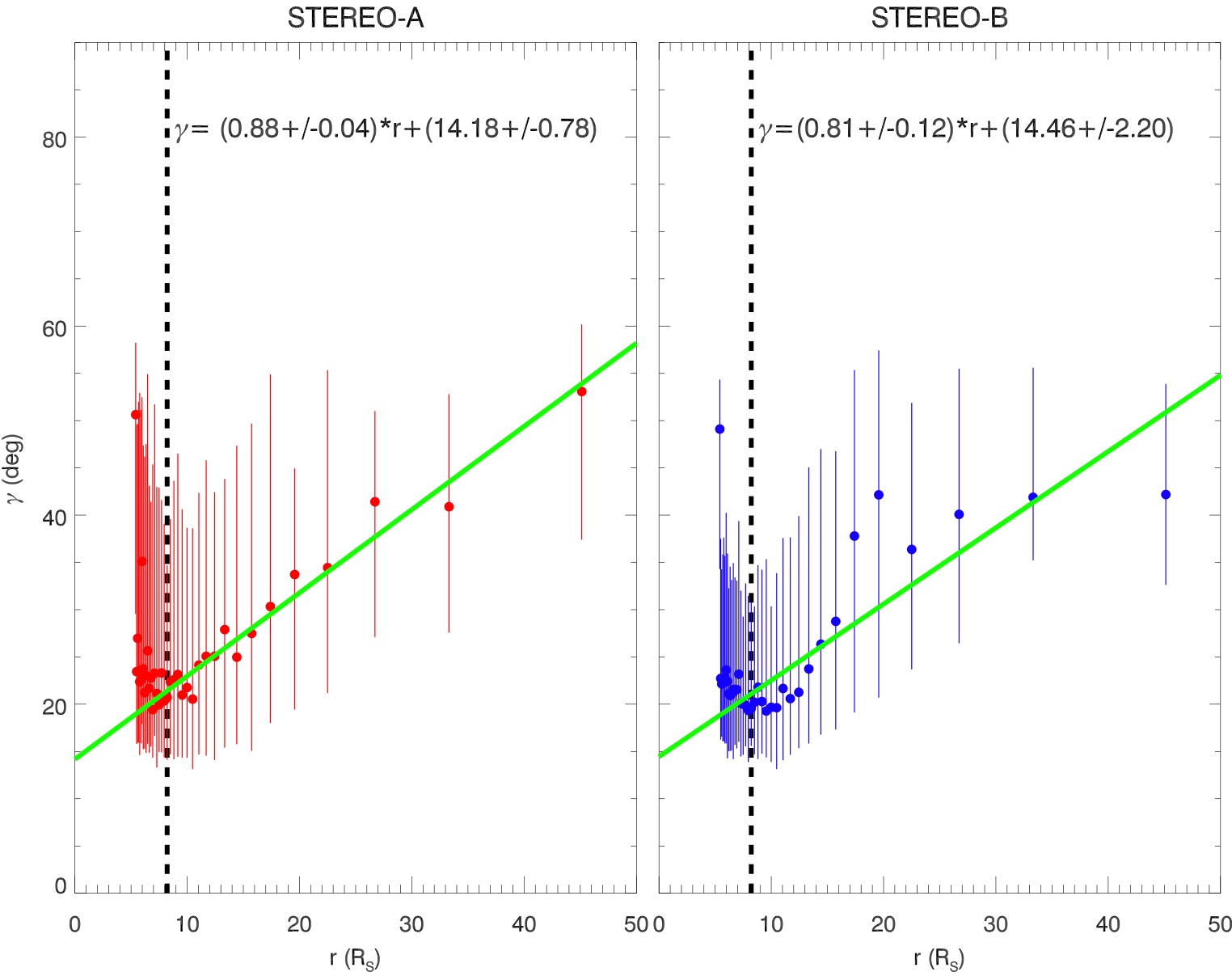}
              }
             \caption{Results of the statistical survey of 152 Type III radio bursts obtained during peak fluxes
							for STEREO-A (on the left) and STEREO-B (on the right). Panels show median values of the apparent source sizes $\gamma$
							as a function of a radial distance from the Sun calculated from the Sittler and Guhathakurta (1999)
							model with an assumption of the fundamental emission. Errorbars represent 25th/75th percentiles.
							Black dashed lines indicate radial distance corresponding to $1$~MHz.
							Green lines are linear fits of points below $1$~MHz. Obtained linear model parameters with one $\sigma$ are located on the top.}
   \label{dist_size}
   \end{figure}

We have used the electron density
model of \citet{1999ApJ...523..812S} to convert frequency channels into radial
distances from the Sun assuming the fundamental emission (Figure \ref{dist_size}).
We fitted a linear model for data points below $1$~MHz
(\textit{i.e.} the statistical maximum flux density)
by minimizing the $\chi^2$ error statistic. Obtained parameters of the model suggest that radio sources expand linearly
as exciter electrons propagate outward from the Sun. \citet{1985A&A...150..205S}
showed that $\gamma$ increases with decreasing frequency just enough to fill a cone with apex located at the Sun
with opening angle of about $80^\circ$ (full width to $1/$e brightness distribution).
We observe the larger opening angle ($140^\circ$) which can be explained by different methods used for
source size calculation.

\subsection{Statistical Results on Radio Source Locations}
\label{S-stat2}

\begin{figure}    
   \centerline{\includegraphics[width=0.9\textwidth,angle=0,clip=]{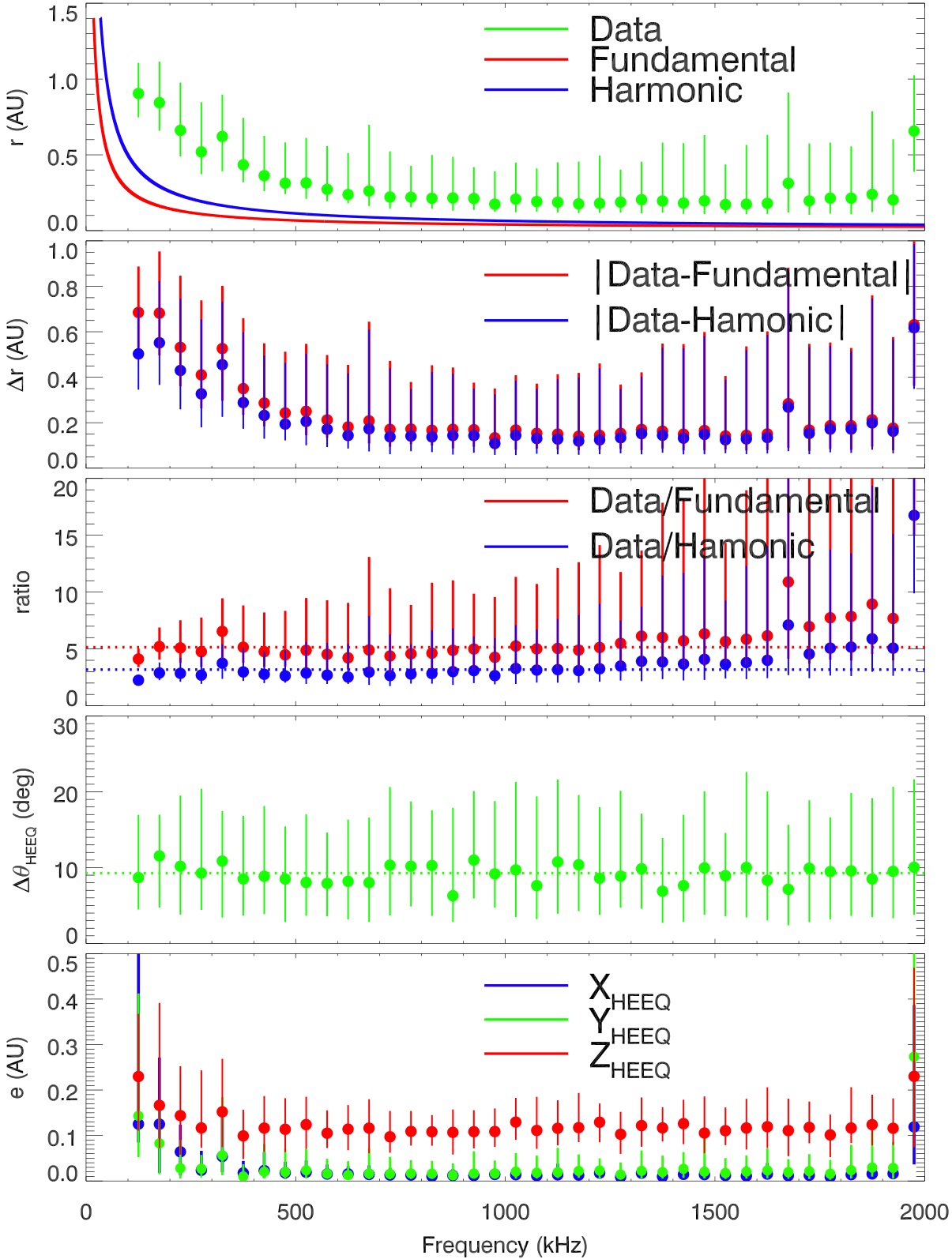}
              }
             \caption{Results of the statistical survey of $98$~Type III bursts observed simultaneously by the two STEREO.
							The first panel contains distances of triangulated radio sources of $98$~Type III bursts vs frequency:
							green circles are median values and error bars represent 25th/75th percentiles. The red and blue lines show
							the theoretical distances using the electron density model of Sittler and Guhathakurta (1999) for the fundamental and harmonic component, respectively.
							The second panel displays absolute differences between observed radio locations and the model.
							The third panel is ratio between observed radio locations and the model. Dotted lines show
							the median values of these ratios.
							The fourth panel displays median values of radio source latitudes calculated in HEEQ.
							The last panel contains medians of uncertainties of the triangulation $e_{\rm{i}}$ [Equation~(15)] in the $X_{\rm{HEEQ}}$, $Y_{\rm{HEEQ}}$, and $Z_{\rm{HEEQ}}$ directions.
							}
   \label{radio_sources}
   \end{figure}

In order to take advantage of two point measurements
we have statistically analyzed Type III radio bursts that have been observed by two spacecraft simultaneously
($98$~events from the total of $152$ discussed in Section \ref{S-stat1}). The topmost panel of Figure \ref{radio_sources}
shows median values of the distance of triangulated radio sources from the Sun \textit{vs.} frequency.
We have used the electron density model of \citet{1999ApJ...523..812S} to assign particular frequencies to
radial distances from the Sun with assumption of the fundamental (F, a red line) and harmonic (H, a blue line) emission.
The highest frequency channel ($1975$~kHz) should not be considered due to HFR switching.

The second panel of Figure \ref{radio_sources} displays absolute differences between the data and both fundamental and harmonic models ($\Delta r$).
This deviation is largest ($\Delta r_{\rm{F}} \approx 0.8$~AU and $\Delta r_{\rm{H}} \approx 0.6$~AU) for the lowest frequency channels.
For frequencies above $\approx 500$~kHz it is roughly constant for both models ($\Delta r_{\rm{F}}$ and $\Delta r_{\rm{H}} \approx 0.2$~AU).
If we compare the source size with a radial diameter of $0.2$~AU,
it corresponds to $\gamma \approx 6^\circ$ when observed from a distance of $1$~AU.

We have compared ratios between the data and model (the third panel of Figure \ref{radio_sources}).
Median values of these ratios are denoted as dotted lines. Statistically, triangulated radio sources are respectively located 
five-times and three-times further from the Sun when compared to the model for the fundamental
and harmonic components. \citet{1984A&A...140...39S} have also observed
radio sources at radial distances from about two-times to five-times further from the Sun
using one spacecraft measurements with assumptions on the IP density profile.

Our results confirm that radio sources lie at considerably larger distances than regions with
electron densities corresponding to particular frequencies \citep{1984A&A...140...39S}.
This suggests, together with extended
apparent source sizes (Figure \ref{stat}), that Type III radio bursts suffer refraction at density gradients and/or
scattering by inhomogeneities. In other words, these propagation effects blur actual locations, polarization
properties, and beaming patterns of the radio sources at kilometric wavelengths
\citep{1985A&A...150..205S,2007ApJ...671..894T}.

In order to investigate magnetic field topology
we calculated latitudes of triangulated radio sources in the HEEQ coordinate system:
\begin{equation}
\theta_{\rm{HEEQ}}=\arctan{\frac{Z_{\rm{HEEQ}}}{\sqrt{X_{\rm{HEEQ}}^2+Y_{\rm{HEEQ}}^2}}}\mbox{.}
\label{eq:lat}
\end{equation}
The third panel of Figure \ref{radio_sources} displays median values of absolute deviations from the solar equatorial
plane ($\Delta\theta_{\rm{HEEQ}}$) \textit{vs.} frequency. Our results suggest that radio sources are statistically confined in this plane
within $10^\circ$ in a given frequency range. Using one spacecraft measurements with assumptions on the
IP density profile \citet{1986ASSL..123..229D} demonstrated that radio sources often start at rather high latitudes,
but then they tend to turn over towards the ecliptic plane at the lower frequencies. 
However, we do not observe this effect in the STEREO data. It might be related to larger uncertainties of radio source locations in the $Z_{\rm{HEEQ}}$ direction.

The fourth panel of Figure \ref{radio_sources} shows uncertainties of the triangulation in the $X_{\rm{HEEQ}}$, $Y_{\rm{HEEQ}}$, and $Z_{\rm{HEEQ}}$
directions calculated as $|\mathbf{r}_{\rm{A}}-\mathbf{r}_{\rm{B}}|$ from Equation~(15).
The obtained values suggest that the largest uncertainties occur in the $Z_{\rm{HEEQ}}$ being roughly $0.1$~AU for all frequencies. 
This can be expected by small separation distances between the spacecraft in this direction. We also observe
increased uncertainties for the highest frequency channel ($1975$~kHz) in all directions which correspond
to the instrumental effects discussed in Section \ref{S-case1}.

\section{Conclusions} 
      \label{S-conclusion}

In this paper we present results of an analysis of solar radio emissions observed by STEREO with a focus on the GP properties.
We have shown two examples of observations of
Type III radio bursts from January 2008 and May 2009. The first event has been already analyzed by \citet{2009SoPh..tmp..100R}
and \citet{2012SoPh..279..153M}. We have achieved a good agreement with abovementioned studies on radio source locations.
The second event occurred in May 2009
when the separation angle between the two STEREO was $\approx90^\circ$ allowing to accurately triangulate radio sources.
Although calculated apparent source sizes are very extended, our results indicate
that electrons responsible for Type III radio bursts propagate along the Parker spiral field lines.
We have also demonstrated that STEREO can be used for stereoscopic investigations
of radio sources even if they are located on the far side of the Sun from
the Earth's perspective.

We have performed a statistical survey of 152 isolated Type III radio bursts observed by STEREO between May 2007 and February 2013.
The maximum flux density $S$ occurs at $\approx 1$~MHz (the first panel of Figure \ref{stat}) and it is discussed in more details
in the linked paper \citep{2014SoPh..289.3121K}.
Statistical results on apparent source sizes $\gamma$ which are very extended for low frequencies ($\gamma \approx 60^\circ$) suggest effects of scattering by density fluctuations in the solar wind.
Absolute values of deviations of wave vectors from the Sun \--- spacecraft line confirm that electrons responsible for Type III radio bursts
statistically propagate in the ecliptic plane as it has been
already observed by \citet{1986ASSL..123..229D}. The 2D degree of polarization is very low.
Triangulated radio sources are statistically located further from the Sun with a factor of $5$ and $3$ for the fundamental
and harmonic components, respectively. This inconsistency has been observed by \citet{1984A&A...140...39S}.
It confirms that we observe only scatter images of real sources.

\section*{Acknowledgements} 
      \label{S-ack}
The authors would like to thank the many individuals and institutions who contributed to making
STEREO/\textit{Waves} possible.
O.~Kruparova thanks the support of the Czech Grant
Agency grant GP13-37174P.
O.~Santolik acknowledges the support of the Czech Grant
Agency grant GAP205/10/2279.
V.~Krupar thanks the support of the Czech Grant
Agency grant GAP209/12/2394.

\end{article} 


\begin{thebibliography}{35}
\ifx\bisbn     \undefined \def\bisbn  #1{ISBN #1}\fi
\ifx\binits    \undefined \def\binits#1{#1}\fi
\ifx\bauthor   \undefined \def\bauthor#1{#1}\fi
\ifx\batitle   \undefined \def\batitle#1{#1}\fi
\ifx\bjtitle   \undefined \def\bjtitle#1{\textit{#1}}\fi
\ifx\bvolume   \undefined \def\bvolume#1{\textbf{#1}}\fi
\ifx\byear     \undefined \def\byear#1{#1}\fi
\ifx\bissue    \undefined \def\bissue#1{#1}\fi
\ifx\bfpage    \undefined \def\bfpage#1{#1}\fi
\ifx\blpage    \undefined \def\blpage #1{#1}\fi
\ifx\burl      \undefined \def\burl#1{\textsf{#1}}\fi
\ifx\href      \undefined \def\href#1#2{\textsf{#2}}\fi
\ifx\betal     \undefined \def\betal{\textit{et al.}}\fi
\ifx\bctitle   \undefined \def\bctitle#1{#1}\fi
\ifx\beditor   \undefined \def\beditor#1{#1}\fi
\ifx\bbtitle   \undefined \def\bbtitle#1{\textit{#1}}\fi
\ifx\bedition  \undefined \def\bedition#1{#1}\fi
\ifx\bseriesno \undefined \def\bseriesno#1{\textbf{#1}}\fi
\ifx\blocation \undefined \def\blocation#1{#1}\fi
\ifx\bsertitle \undefined \def\bsertitle#1{\textit{#1}}\fi
\ifx\bsnm      \undefined \def\bsnm#1{#1}\fi
\ifx\bsuffix   \undefined \def\bsuffix#1{#1}\fi
\ifx\bparticle \undefined \def\bparticle#1{#1}\fi
\ifx\barticle  \undefined \def\barticle#1{}\fi
\ifx\binstitute  \undefined \def\binstitute#1{#1}\fi
\ifx\bpublisher  \undefined \def\bpublisher#1{#1}\fi
\ifx\doiurl    \undefined
  \def\doiurl#1{\href{http://dx.doi.org/#1}{\textsf{DOI}}}\fi
\ifx\arxivurl  \undefined
  \def\arxivurl#1{\href{http://arxiv.org/abs/#1}{\textsf{arXiv}}}\fi
\ifx\adsurl    \undefined
  \def\adsurl#1{\href{http://adsabs.harvard.edu/abs/#1}{\textsf{ADS}}}\fi
\ifx\botherref \undefined \def\botherref#1{}\fi
\ifx\url       \undefined \def\url#1{\textsf{#1}}\fi
\ifx\bchapter  \undefined \def\bchapter#1{}\fi
\ifx\bbook     \undefined \def\bbook#1{}\fi
\ifx\bcomment  \undefined \def\bcomment#1{#1}\fi
\ifx\oauthor   \undefined \def\oauthor#1{#1}\fi
\ifx\citeauthoryear \undefined\def \citeauthoryear#1{#1}\fi
\def\endbibitem {}
\ifx\bconflocation  \undefined \def\bconflocation#1{#1} \fi

\bibitem[\protect\citeauthoryear{{Bale}
  \textit{et~al.}}{2008}]{2008SSRv..136..529B}
\begin{barticle}
\bauthor{\bsnm{{Bale}}, \binits{S.D.}},
\bauthor{\bsnm{{Ullrich}}, \binits{R.}},
\bauthor{\bsnm{{Goetz}}, \binits{K.}},
\bauthor{\bsnm{{Alster}}, \binits{N.}},
\bauthor{\bsnm{{Cecconi}}, \binits{B.}},
\bauthor{\bsnm{{Dekkali}}, \binits{M.}},
\bauthor{\bsnm{{Lingner}}, \binits{N.R.}},
\bauthor{\bsnm{{Macher}}, \binits{W.}},
\bauthor{\bsnm{{Manning}}, \binits{R.E.}},
\bauthor{\bsnm{{McCauley}}, \binits{J.}},
\bauthor{\bsnm{{Monson}}, \binits{S.J.}},
\bauthor{\bsnm{{Oswald}}, \binits{T.H.}},
\bauthor{\bsnm{{Pulupa}}, \binits{M.}}:
\byear{2008},
\batitle{{The electric antennas for the STEREO/WAVES experiment}}.
\bjtitle{Space Sci. Rev.}
\bvolume{136},
\bfpage{529}.
\doiurl{10.1007/s11214-007-9251-x}.
\adsurl{http://cdsads.u-strasbg.fr/abs/2008SSRv..136..529B}.
\end{barticle}
\endbibitem

\bibitem[\protect\citeauthoryear{{Bougeret}, {Fainberg}, and
  {Stone}}{1984}]{1984A&A...141...17B}
\begin{barticle}
\bauthor{\bsnm{{Bougeret}}, \binits{J.-L.}},
\bauthor{\bsnm{{Fainberg}}, \binits{J.}},
\bauthor{\bsnm{{Stone}}, \binits{R.G.}}:
\byear{1984},
\batitle{{Interplanetary radio storms. II - Emission levels and solar wind
  speed in the range 0.05-0.8 AU}}.
\bjtitle{\aap}
\bvolume{141},
\bfpage{17}.
\adsurl{1984A\%26A...141...17B}.
\end{barticle}
\endbibitem

\bibitem[\protect\citeauthoryear{{Bougeret}
  \textit{et~al.}}{2008}]{2008SSRv..136..487B}
\begin{barticle}
\bauthor{\bsnm{{Bougeret}}, \binits{J.L.}},
\bauthor{\bsnm{{Goetz}}, \binits{K.}},
\bauthor{\bsnm{{Kaiser}}, \binits{M.L.}},
\bauthor{\bsnm{{Bale}}, \binits{S.D.}},
\bauthor{\bsnm{{Kellogg}}, \binits{P.J.}},
\bauthor{\bsnm{{Maksimovic}}, \binits{M.}},
\bauthor{\bsnm{{Monge}}, \binits{N.}},
\bauthor{\bsnm{{Monson}}, \binits{S.J.}},
\bauthor{\bsnm{{Astier}}, \binits{P.L.}},
\bauthor{\bsnm{{Davy}}, \binits{S.}},
\bauthor{\bsnm{{Dekkali}}, \binits{M.}},
\bauthor{\bsnm{{Hinze}}, \binits{J.J.}},
\bauthor{\bsnm{{Manning}}, \binits{R.E.}},
\bauthor{\bsnm{{Aguilar-Rodriguez}}, \binits{E.}},
\bauthor{\bsnm{{Bonnin}}, \binits{X.}},
\bauthor{\bsnm{{Briand}}, \binits{C.}},
\bauthor{\bsnm{{Cairns}}, \binits{I.H.}},
\bauthor{\bsnm{{Cattell}}, \binits{C.A.}},
\bauthor{\bsnm{{Cecconi}}, \binits{B.}},
\bauthor{\bsnm{{Eastwood}}, \binits{J.}},
\bauthor{\bsnm{{Ergun}}, \binits{R.E.}},
\bauthor{\bsnm{{Fainberg}}, \binits{J.}},
\bauthor{\bsnm{{Hoang}}, \binits{S.}},
\bauthor{\bsnm{{Huttunen}}, \binits{K.E.J.}},
\bauthor{\bsnm{{Krucker}}, \binits{S.}},
\bauthor{\bsnm{{Lecacheux}}, \binits{A.}},
\bauthor{\bsnm{{MacDowall}}, \binits{R.J.}},
\bauthor{\bsnm{{Macher}}, \binits{W.}},
\bauthor{\bsnm{{Mangeney}}, \binits{A.}},
\bauthor{\bsnm{{Meetre}}, \binits{C.A.}},
\bauthor{\bsnm{{Moussas}}, \binits{X.}},
\bauthor{\bsnm{{Nguyen}}, \binits{Q.N.}},
\bauthor{\bsnm{{Oswald}}, \binits{T.H.}},
\bauthor{\bsnm{{Pulupa}}, \binits{M.}},
\bauthor{\bsnm{{Reiner}}, \binits{M.J.}},
\bauthor{\bsnm{{Robinson}}, \binits{P.A.}},
\bauthor{\bsnm{{Rucker}}, \binits{H.}},
\bauthor{\bsnm{{Salem}}, \binits{C.}},
\bauthor{\bsnm{{Santolik}}, \binits{O.}},
\bauthor{\bsnm{{Silvis}}, \binits{J.M.}},
\bauthor{\bsnm{{Ullrich}}, \binits{R.}},
\bauthor{\bsnm{{Zarka}}, \binits{P.}},
\bauthor{\bsnm{{Zouganelis}}, \binits{I.}}:
\byear{2008},
\batitle{{S/WAVES: The radio and plasma wave investigation on the STEREO
  mission}}.
\bjtitle{Space Sci. Rev.}
\bvolume{136},
\bfpage{487}.
\doiurl{10.1007/s11214-007-9298-8}.
\adsurl{http://cdsads.u-strasbg.fr/abs/2008SSRv..136..487B}.
\end{barticle}
\endbibitem

\bibitem[\protect\citeauthoryear{{Cecconi}
  \textit{et~al.}}{2008}]{2008SSRv..136..549C}
\begin{barticle}
\bauthor{\bsnm{{Cecconi}}, \binits{B.}},
\bauthor{\bsnm{{Bonnin}}, \binits{X.}},
\bauthor{\bsnm{{Hoang}}, \binits{S.}},
\bauthor{\bsnm{{Maksimovic}}, \binits{M.}},
\bauthor{\bsnm{{Bale}}, \binits{S.D.}},
\bauthor{\bsnm{{Bougeret}}, \binits{J.-L.}},
\bauthor{\bsnm{{Goetz}}, \binits{K.}},
\bauthor{\bsnm{{Lecacheux}}, \binits{A.}},
\bauthor{\bsnm{{Reiner}}, \binits{M.J.}},
\bauthor{\bsnm{{Rucker}}, \binits{H.O.}},
\bauthor{\bsnm{{Zarka}}, \binits{P.}}:
\byear{2008},
\batitle{{STEREO/Waves goniopolarimetry}}.
\bjtitle{Space Sci. Rev.}
\bvolume{136},
\bfpage{549}.
\doiurl{10.1007/s11214-007-9255-6}.
\adsurl{2008SSRv..136..549C}.
\end{barticle}
\endbibitem

\bibitem[\protect\citeauthoryear{{Dulk} and
  {Suzuki}}{1980}]{1980A&A....88..203D}
\begin{barticle}
\bauthor{\bsnm{{Dulk}}, \binits{G.A.}},
\bauthor{\bsnm{{Suzuki}}, \binits{S.}}:
\byear{1980},
\batitle{{The position and polarization of type III solar bursts}}.
\bjtitle{Astron. and Astrophys.}
\bvolume{88},
\bfpage{203}.
\adsurl{1980A\%26A....88..203D}.
\end{barticle}
\endbibitem

\bibitem[\protect\citeauthoryear{{Dulk}, {Steinberg}, and
  {Hoang}}{1984}]{1984A&A...141...30D}
\begin{barticle}
\bauthor{\bsnm{{Dulk}}, \binits{G.A.}},
\bauthor{\bsnm{{Steinberg}}, \binits{J.L.}},
\bauthor{\bsnm{{Hoang}}, \binits{S.}}:
\byear{1984},
\batitle{{Type III bursts in interplanetary space - Fundamental or harmonic?}}
\bjtitle{\aap}
\bvolume{141},
\bfpage{30}.
\adsurl{1984A\%26A...141...30D}.
\end{barticle}
\endbibitem

\bibitem[\protect\citeauthoryear{{Dulk}
  \textit{et~al.}}{1986}]{1986ASSL..123..229D}
\begin{bchapter}
\bauthor{\bsnm{{Dulk}}, \binits{G.A.}},
\bauthor{\bsnm{{Steinberg}}, \binits{J.L.}},
\bauthor{\bsnm{{Hoang}}, \binits{S.}},
\bauthor{\bsnm{{Lecacheux}}, \binits{A.}}:
\byear{1986},
\bctitle{{Latitude distribution of interplanetary magnetic field lines rooted
  in active regions}}.
In: \beditor{\bsnm{{Marsden}}, \binits{R.G.}} (ed.)
\bbtitle{The Sun and the Heliosphere in Three Dimensions},
\bsertitle{D. Reidel Publishing Co., Dordrecht}
\bseriesno{123},
\bfpage{229}.
\adsurl{1986ASSL..123..229D}.
\end{bchapter}
\endbibitem

\bibitem[\protect\citeauthoryear{{Dulk}
  \textit{et~al.}}{1987}]{1987A&A...173..366D}
\begin{barticle}
\bauthor{\bsnm{{Dulk}}, \binits{G.A.}},
\bauthor{\bsnm{{Goldman}}, \binits{M.V.}},
\bauthor{\bsnm{{Steinberg}}, \binits{J.L.}},
\bauthor{\bsnm{{Hoang}}, \binits{S.}}:
\byear{1987},
\batitle{{The speeds of electrons that excite solar radio bursts of type III}}.
\bjtitle{Astron. and Astrophys.}
\bvolume{173},
\bfpage{366}.
\adsurl{1987A\%26A...173..366D}.
\end{barticle}
\endbibitem

\bibitem[\protect\citeauthoryear{{Fainberg}, {Evans}, and
  {Stone}}{1972}]{1972Sci...178..743F}
\begin{barticle}
\bauthor{\bsnm{{Fainberg}}, \binits{J.}},
\bauthor{\bsnm{{Evans}}, \binits{L.G.}},
\bauthor{\bsnm{{Stone}}, \binits{R.G.}}:
\byear{1972},
\batitle{{Radio tracking of solar energetic particles through interplanetary
  space}}.
\bjtitle{Science}
\bvolume{178},
\bfpage{743}.
\doiurl{10.1126/science.178.4062.743}.
\adsurl{1972Sci...178..743F}.
\end{barticle}
\endbibitem

\bibitem[\protect\citeauthoryear{{Fitzenreiter}
  \textit{et~al.}}{1977}]{1977SoPh...52..477F}
\begin{barticle}
\bauthor{\bsnm{{Fitzenreiter}}, \binits{R.J.}},
\bauthor{\bsnm{{Fainberg}}, \binits{J.}},
\bauthor{\bsnm{{Weber}}, \binits{R.R.}},
\bauthor{\bsnm{{Alvarez}}, \binits{H.}},
\bauthor{\bsnm{{Haddock}}, \binits{F.T.}},
\bauthor{\bsnm{{Potter}}, \binits{W.H.}}:
\byear{1977},
\batitle{{Radio observations of interplanetary magnetic field structures out of
  the ecliptic}}.
\bjtitle{\solphys}
\bvolume{52},
\bfpage{477}.
\doiurl{10.1007/BF00149662}.
\adsurl{1977SoPh...52..477F}.
\end{barticle}
\endbibitem

\bibitem[\protect\citeauthoryear{{Ginzburg} and
  {Zhelezniakov}}{1958}]{1958SvA.....2..653G}
\begin{barticle}
\bauthor{\bsnm{{Ginzburg}}, \binits{V.L.}},
\bauthor{\bsnm{{Zhelezniakov}}, \binits{V.V.}}:
\byear{1958},
\batitle{{On the possible mechanisms of sporadic solar radio emission
  (Radiation in an isotropic plasma)}}.
\bjtitle{Soviet Astron.}
\bvolume{2},
\bfpage{653}.
\adsurl{1958SvA.....2..653G}.
\end{barticle}
\endbibitem

\bibitem[\protect\citeauthoryear{{Gurnett} and
  {Anderson}}{1976}]{1976Sci...194.1159G}
\begin{barticle}
\bauthor{\bsnm{{Gurnett}}, \binits{D.A.}},
\bauthor{\bsnm{{Anderson}}, \binits{R.R.}}:
\byear{1976},
\batitle{{Electron plasma oscillations associated with type III radio bursts}}.
\bjtitle{Science}
\bvolume{194},
\bfpage{1159}.
\doiurl{10.1126/science.194.4270.1159}.
\adsurl{1976Sci...194.1159G}.
\end{barticle}
\endbibitem

\bibitem[\protect\citeauthoryear{{Hoang}
  \textit{et~al.}}{1981}]{1981JGR....86.4531H}
\begin{barticle}
\bauthor{\bsnm{{Hoang}}, \binits{S.}},
\bauthor{\bsnm{{Steinberg}}, \binits{J.L.}},
\bauthor{\bsnm{{Stone}}, \binits{R.G.}},
\bauthor{\bsnm{{Zwickl}}, \binits{R.H.}},
\bauthor{\bsnm{{Fainberg}}, \binits{J.}}:
\byear{1981},
\batitle{{The 2f/p/ circumterrestrial radio radiation as seen from ISEE 3}}.
\bjtitle{J. Geophys. Res.}
\bvolume{86},
\bfpage{4531}.
\doiurl{10.1029/JA086iA06p04531}.
\adsurl{1981JGR....86.4531H}.
\end{barticle}
\endbibitem

\bibitem[\protect\citeauthoryear{{Kaiser}
  \textit{et~al.}}{2008}]{2008SSRv..136....5K}
\begin{barticle}
\bauthor{\bsnm{{Kaiser}}, \binits{M.L.}},
\bauthor{\bsnm{{Kucera}}, \binits{T.A.}},
\bauthor{\bsnm{{Davila}}, \binits{J.M.}},
\bauthor{\bsnm{{St.~Cyr}}, \binits{O.C.}},
\bauthor{\bsnm{{Guhathakurta}}, \binits{M.}},
\bauthor{\bsnm{{Christian}}, \binits{E.}}:
\byear{2008},
\batitle{{The STEREO mission: An introduction}}.
\bjtitle{Space Sci. Rev.}
\bvolume{136},
\bfpage{5}.
\doiurl{10.1007/s11214-007-9277-0}.
\adsurl{2008SSRv..136....5K}.
\end{barticle}
\endbibitem

\bibitem[\protect\citeauthoryear{{Krupar}
  \textit{et~al.}}{2010}]{2010AIPC.1216..284K}
\begin{bchapter}
\bauthor{\bsnm{{Krupar}}, \binits{V.}},
\bauthor{\bsnm{{Maksimovic}}, \binits{M.}},
\bauthor{\bsnm{{Santolik}}, \binits{O.}},
\bauthor{\bsnm{{Cecconi}}, \binits{B.}},
\bauthor{\bsnm{{Nguyen}}, \binits{Q.N.}},
\bauthor{\bsnm{{Hoang}}, \binits{S.}},
\bauthor{\bsnm{{Goetz}}, \binits{K.}}:
\byear{2010},
\bctitle{{The apparent source size of type III radio bursts: Preliminary
  results by the STEREO/WAVES instruments}}.
In: \beditor{\bsnm{{Maksimovic}}, \binits{M.}},
\beditor{\bsnm{{Issautier}}, \binits{K.}},
\beditor{\bsnm{{Meyer-Vernet}}, \binits{N.}},
\beditor{\bsnm{{Moncuquet}}, \binits{M.}},
\beditor{\bsnm{{Pantellini}}, \binits{F.}} (eds.)
\bbtitle{AIP Conf. Ser.}
\bseriesno{1216},
\bfpage{284}.
\doiurl{10.1063/1.3395856}.
\adsurl{2010AIPC.1216..284K}.
\end{bchapter}
\endbibitem

\bibitem[\protect\citeauthoryear{{Krupar}
  \textit{et~al.}}{2012}]{2012JGRA..11706101K}
\begin{barticle}
\bauthor{\bsnm{{Krupar}}, \binits{V.}},
\bauthor{\bsnm{{Santolik}}, \binits{O.}},
\bauthor{\bsnm{{Cecconi}}, \binits{B.}},
\bauthor{\bsnm{{Maksimovic}}, \binits{M.}},
\bauthor{\bsnm{{Bonnin}}, \binits{X.}},
\bauthor{\bsnm{{Panchenko}}, \binits{M.}},
\bauthor{\bsnm{{Zaslavsky}}, \binits{A.}}:
\byear{2012},
\batitle{{Goniopolarimetric inversion using SVD: An application to type III
  radio bursts observed by STEREO}}.
\bjtitle{J. Geophys. Res.}
\bvolume{117},
\bfpage{6101}.
\doiurl{10.1029/2011JA017333}.
\adsurl{2012JGRA..11706101K}.
\end{barticle}
\endbibitem

\bibitem[\protect\citeauthoryear{{Krupar}
  \textit{et~al.}}{2014}]{2014SoPh..289.3121K}
\begin{barticle}
\bauthor{\bsnm{{Krupar}}, \binits{V.}},
\bauthor{\bsnm{{Maksimovic}}, \binits{M.}},
\bauthor{\bsnm{{Santolik}}, \binits{O.}},
\bauthor{\bsnm{{Kontar}}, \binits{E.P.}},
\bauthor{\bsnm{{Cecconi}}, \binits{B.}},
\bauthor{\bsnm{{Hoang}}, \binits{S.}},
\bauthor{\bsnm{{Kruparova}}, \binits{O.}},
\bauthor{\bsnm{{Soucek}}, \binits{J.}},
\bauthor{\bsnm{{Reid}}, \binits{H.}},
\bauthor{\bsnm{{Zaslavsky}}, \binits{A.}}:
\byear{2014},
\batitle{{Statistical survey of type III radio bursts at long wavelengths
  observed by the Solar TErrestrial RElations Observatory (STEREO)/ Waves
  instruments: Radio flux density variations with frequency}}.
\bjtitle{\solphys}
\bvolume{289},
\bfpage{3121}.
\doiurl{10.1007/s11207-014-0522-x}.
\adsurl{2014SoPh..289.3121K}.
\end{barticle}
\endbibitem

\bibitem[\protect\citeauthoryear{{Lin}
  \textit{et~al.}}{1981}]{1981ApJ...251..364L}
\begin{barticle}
\bauthor{\bsnm{{Lin}}, \binits{R.P.}},
\bauthor{\bsnm{{Potter}}, \binits{D.W.}},
\bauthor{\bsnm{{Gurnett}}, \binits{D.A.}},
\bauthor{\bsnm{{Scarf}}, \binits{F.L.}}:
\byear{1981},
\batitle{{Energetic electrons and plasma waves associated with a solar type III
  radio burst}}.
\bjtitle{Astrophys. J.}
\bvolume{251},
\bfpage{364}.
\doiurl{10.1086/159471}.
\adsurl{1981ApJ...251..364L}.
\end{barticle}
\endbibitem

\bibitem[\protect\citeauthoryear{{Manning} and
  {Fainberg}}{1980}]{1980SSI.....5..161M}
\begin{barticle}
\bauthor{\bsnm{{Manning}}, \binits{R.}},
\bauthor{\bsnm{{Fainberg}}, \binits{J.}}:
\byear{1980},
\batitle{{A new method of measuring radio source parameters of a partially
  polarized distributed source from spacecraft observations}}.
\bjtitle{Space Sci. Instrum.}
\bvolume{5},
\bfpage{161}.
\adsurl{1980SSI.....5..161M}.
\end{barticle}
\endbibitem

\bibitem[\protect\citeauthoryear{{Martens} and
  {Kuin}}{1989}]{1989SoPh..122..263M}
\begin{barticle}
\bauthor{\bsnm{{Martens}}, \binits{P.C.H.}},
\bauthor{\bsnm{{Kuin}}, \binits{N.P.M.}}:
\byear{1989},
\batitle{{A circuit model for filament eruptions and two-ribbon flares}}.
\bjtitle{Solar Phys.}
\bvolume{122},
\bfpage{263}.
\doiurl{10.1007/BF00912996}.
\adsurl{1989SoPh..122..263M}.
\end{barticle}
\endbibitem

\bibitem[\protect\citeauthoryear{{Mart{\'{\i}}nez-Oliveros}
  \textit{et~al.}}{2012}]{2012SoPh..279..153M}
\begin{barticle}
\bauthor{\bsnm{{Mart{\'{\i}}nez-Oliveros}}, \binits{J.C.}},
\bauthor{\bsnm{{Lindsey}}, \binits{C.}},
\bauthor{\bsnm{{Bale}}, \binits{S.D.}},
\bauthor{\bsnm{{Krucker}}, \binits{S.}}:
\byear{2012},
\batitle{{Determination of electromagnetic source direction as an eigenvalue
  problem}}.
\bjtitle{Solar Phys.}
\bvolume{279},
\bfpage{153}.
\doiurl{10.1007/s11207-012-9998-4}.
\adsurl{2012SoPh..279..153M}.
\end{barticle}
\endbibitem

\bibitem[\protect\citeauthoryear{{Melrose}}{1980}]{1980SSRv...26....3M}
\begin{barticle}
\bauthor{\bsnm{{Melrose}}, \binits{D.B.}}:
\byear{1980},
\batitle{{The emission mechanisms for solar radio bursts}}.
\bjtitle{Space Sci. Rev.}
\bvolume{26},
\bfpage{3}.
\doiurl{10.1007/BF00212597}.
\adsurl{1980SSRv...26....3M}.
\end{barticle}
\endbibitem

\bibitem[\protect\citeauthoryear{{Parker}}{1958}]{1958ApJ...128..664P}
\begin{barticle}
\bauthor{\bsnm{{Parker}}, \binits{E.N.}}:
\byear{1958},
\batitle{{Dynamics of the interplanetary gas and magnetic fields.}}
\bjtitle{\apj}
\bvolume{128},
\bfpage{664}.
\doiurl{10.1086/146579}.
\adsurl{1958ApJ...128..664P}.
\end{barticle}
\endbibitem

\bibitem[\protect\citeauthoryear{{Reiner}
  \textit{et~al.}}{1998}]{1998JGR...103.1923R}
\begin{barticle}
\bauthor{\bsnm{{Reiner}}, \binits{M.J.}},
\bauthor{\bsnm{{Fainberg}}, \binits{J.}},
\bauthor{\bsnm{{Kaiser}}, \binits{M.L.}},
\bauthor{\bsnm{{Stone}}, \binits{R.G.}}:
\byear{1998},
\batitle{{Type III radio source located by Ulysses/Wind triangulation}}.
\bjtitle{\jgr}
\bvolume{103},
\bfpage{1923}.
\doiurl{10.1029/97JA02646}.
\adsurl{1998JGR...103.1923R}.
\end{barticle}
\endbibitem

\bibitem[\protect\citeauthoryear{{Reiner}
  \textit{et~al.}}{2009}]{2009SoPh..tmp..100R}
\begin{botherref}
\oauthor{\bsnm{{Reiner}}, \binits{M.J.}},
\oauthor{\bsnm{{Goetz}}, \binits{K.}},
\oauthor{\bsnm{{Fainberg}}, \binits{J.}},
\oauthor{\bsnm{{Kaiser}}, \binits{M.L.}},
\oauthor{\bsnm{{Maksimovic}}, \binits{M.}},
\oauthor{\bsnm{{Cecconi}}, \binits{B.}},
\oauthor{\bsnm{{Hoang}}, \binits{S.}},
\oauthor{\bsnm{{Bale}}, \binits{S.D.}},
\oauthor{\bsnm{{Bougeret}}, \binits{J.-L.}}:
2009,
{Multipoint observations of solar type III radio bursts from STEREO and Wind}.
\textit{Solar Phys.},
100.
\doiurl{10.1007/s11207-009-9404-z}.
\adsurl{2009SoPh..tmp..100R}.
\end{botherref}
\endbibitem

\bibitem[\protect\citeauthoryear{{Santol{\'{\i}}k}, {Parrot}, and
  {Lefeuvre}}{2003}]{2003RaSc...38a..10S}
\begin{barticle}
\bauthor{\bsnm{{Santol{\'{\i}}k}}, \binits{O.}},
\bauthor{\bsnm{{Parrot}}, \binits{M.}},
\bauthor{\bsnm{{Lefeuvre}}, \binits{F.}}:
\byear{2003},
\batitle{{Singular value decomposition methods for wave propagation analysis}}.
\bjtitle{Radio Sci.}
\bvolume{38},
\bfpage{010000}.
\doiurl{10.1029/2000RS002523}.
\adsurl{2003RaSc...38a..10S}.
\end{barticle}
\endbibitem

\bibitem[\protect\citeauthoryear{{Santol{\'{\i}}k}
  \textit{et~al.}}{2002}]{2002JGRA..107.1444S}
\begin{barticle}
\bauthor{\bsnm{{Santol{\'{\i}}k}}, \binits{O.}},
\bauthor{\bsnm{{Pickett}}, \binits{J.S.}},
\bauthor{\bsnm{{Gurnett}}, \binits{D.A.}},
\bauthor{\bsnm{{Storey}}, \binits{L.R.O.}}:
\byear{2002},
\batitle{{Magnetic component of narrowband ion cyclotron waves in the auroral
  zone}}.
\bjtitle{J. Geophys. Res.}
\bvolume{107},
\bfpage{1444}.
\doiurl{10.1029/2001JA000146}.
\adsurl{2002JGRA..107.1444S}.
\end{barticle}
\endbibitem

\bibitem[\protect\citeauthoryear{{Sittler} and
  {Guhathakurta}}{1999}]{1999ApJ...523..812S}
\begin{barticle}
\bauthor{\bsnm{{Sittler}}, \binits{E.C.} \bsuffix{Jr.}},
\bauthor{\bsnm{{Guhathakurta}}, \binits{M.}}:
\byear{1999},
\batitle{{Semiempirical two-dimensional magnetohydrodynamic model of the solar
  corona and interplanetary medium}}.
\bjtitle{Astrophys. J.}
\bvolume{523},
\bfpage{812}.
\doiurl{10.1086/307742}.
\adsurl{1999ApJ...523..812S}.
\end{barticle}
\endbibitem

\bibitem[\protect\citeauthoryear{{Steinberg}, {Hoang}, and
  {Dulk}}{1985}]{1985A&A...150..205S}
\begin{barticle}
\bauthor{\bsnm{{Steinberg}}, \binits{J.L.}},
\bauthor{\bsnm{{Hoang}}, \binits{S.}},
\bauthor{\bsnm{{Dulk}}, \binits{G.A.}}:
\byear{1985},
\batitle{{Evidence of scattering effects on the sizes of interplanetary type
  III radio bursts}}.
\bjtitle{Astron. and Astrophys.}
\bvolume{150},
\bfpage{205}.
\adsurl{1985A\%26A...150..205S}.
\end{barticle}
\endbibitem

\bibitem[\protect\citeauthoryear{{Steinberg}
  \textit{et~al.}}{1984}]{1984A&A...140...39S}
\begin{barticle}
\bauthor{\bsnm{{Steinberg}}, \binits{J.L.}},
\bauthor{\bsnm{{Dulk}}, \binits{G.A.}},
\bauthor{\bsnm{{Hoang}}, \binits{S.}},
\bauthor{\bsnm{{Lecacheux}}, \binits{A.}},
\bauthor{\bsnm{{Aubier}}, \binits{M.G.}}:
\byear{1984},
\batitle{{Type III radio bursts in the interplanetary medium - The role of
  propagation}}.
\bjtitle{Astron. and Astrophys.}
\bvolume{140},
\bfpage{39}.
\adsurl{1984A\%26A...140...39S}.
\end{barticle}
\endbibitem

\bibitem[\protect\citeauthoryear{{Thejappa}, {MacDowall}, and
  {Kaiser}}{2007}]{2007ApJ...671..894T}
\begin{barticle}
\bauthor{\bsnm{{Thejappa}}, \binits{G.}},
\bauthor{\bsnm{{MacDowall}}, \binits{R.J.}},
\bauthor{\bsnm{{Kaiser}}, \binits{M.L.}}:
\byear{2007},
\batitle{{Monte Carlo simulation of directivity of interplanetary radio
  bursts}}.
\bjtitle{Astrophys. J.}
\bvolume{671},
\bfpage{894}.
\doiurl{10.1086/522664}.
\adsurl{2007ApJ...671..894T}.
\end{barticle}
\endbibitem

\bibitem[\protect\citeauthoryear{{Weber}}{1978}]{1978SoPh...59..377W}
\begin{barticle}
\bauthor{\bsnm{{Weber}}, \binits{R.R.}}:
\byear{1978},
\batitle{{Low frequency spectra of type III solar radio bursts}}.
\bjtitle{Solar Phys.}
\bvolume{59},
\bfpage{377}.
\doiurl{10.1007/BF00951843}.
\adsurl{1978SoPh...59..377W}.
\end{barticle}
\endbibitem

\bibitem[\protect\citeauthoryear{{Weber}
  \textit{et~al.}}{1977}]{1977SoPh...54..431W}
\begin{barticle}
\bauthor{\bsnm{{Weber}}, \binits{R.R.}},
\bauthor{\bsnm{{Fitzenreiter}}, \binits{R.J.}},
\bauthor{\bsnm{{Novaco}}, \binits{J.C.}},
\bauthor{\bsnm{{Fainberg}}, \binits{J.}}:
\byear{1977},
\batitle{{Interplanetary baseline observations of type III solar radio
  bursts}}.
\bjtitle{\solphys}
\bvolume{54},
\bfpage{431}.
\doiurl{10.1007/BF00159934}.
\adsurl{1977SoPh...54..431W}.
\end{barticle}
\endbibitem

\bibitem[\protect\citeauthoryear{{Wild}}{1950}]{1950AuSRA...3..541W}
\begin{barticle}
\bauthor{\bsnm{{Wild}}, \binits{J.P.}}:
\byear{1950},
\batitle{{Observations of the spectrum of high-intensity solar radiation at
  metre wavelengths. III. Isolated bursts}}.
\bjtitle{Aust. J. Sci. Res. A}
\bvolume{3},
\bfpage{541}.
\adsurl{1950AuSRA...3..541W}.
\end{barticle}
\endbibitem

\bibitem[\protect\citeauthoryear{{Zaslavsky}
  \textit{et~al.}}{2011}]{2011RaSc...46.2008Z}
\begin{barticle}
\bauthor{\bsnm{{Zaslavsky}}, \binits{A.}},
\bauthor{\bsnm{{Meyer-Vernet}}, \binits{N.}},
\bauthor{\bsnm{{Hoang}}, \binits{S.}},
\bauthor{\bsnm{{Maksimovic}}, \binits{M.}},
\bauthor{\bsnm{{Bale}}, \binits{S.D.}}:
\byear{2011},
\batitle{{On the antenna calibration of space radio instruments using the
  galactic background: General formulas and application to STEREO/WAVES}}.
\bjtitle{Radio Sci.}
\bvolume{46},
\bfpage{RS2008}.
\doiurl{10.1029/2010RS004464}.
\adsurl{2011RaSc...46.2008Z}.
\end{barticle}
\endbibitem

\end{thebibliography}
\end{document}